\documentclass[acmsmall,screen,nonacm]{acmart}

\usepackage{url}
\usepackage{wrapfig}
\usepackage{algorithmic}
\usepackage{graphicx}
\usepackage{subcaption}
\usepackage[export]{adjustbox}
\usepackage{textcomp}
\usepackage{booktabs}
\usepackage{siunitx}
\usepackage{xcolor}
\usepackage{makecell}
\usepackage{listings}
\usepackage{color}
\usepackage{multirow}
\usepackage{tabularx}
\usepackage{colortbl}
\usepackage{bm}
\usepackage{multicol}
\usepackage[ruled,vlined, linesnumbered]{algorithm2e}
\usepackage{amsmath}
\usepackage{mdframed}
\usepackage{enumitem}
\usepackage{amsmath}
\usepackage{float}
\usepackage[absolute,overlay]{textpos}
\usepackage{tabularx}
\usepackage[utf8]{inputenc}
\usepackage{booktabs}
\usepackage{wrapfig,placeins}

\makeatletter
\newcommand{\removelatexcentering}{\setlength{\ALG@thistlm}{\leftmargin}}
\makeatother

\lstdefinelanguage{yaml}{
  keywords={true,false,null,y,n},
  keywordstyle=\color{darkgray}\bfseries,
  sensitive=false,
  comment=[l]{\#},
  commentstyle=\color{purple}\ttfamily,
  stringstyle=\color{blue}\ttfamily,
  morestring=[b]',
  morestring=[b]"
}

\lstdefinelanguage{python}{
   keywords={sem, import, node, ignore, take, entry, activity, exit, spawn, 
           edge, walker, or, if, elif, else, for,  by, while,
           continue, break, disengage, report, anchor, has, can, true, false, from,
           context, info, try, strict, length, test, type, str, Optional,
           int, float, list, dict, bool, digraph, subgraph, test, by, 
           skip, assert, yield, class, def, Method, Function, pass,except,strip,split,
       },
   sensitive=false, 
   morecomment=[l]{\#}, 
   morecomment=[s]{```}{```}, 
   moredelim=[is][\color{blue}]{\{}{\}}, 
   morestring=[b]", 
   morestring=[b]`,
   morestring=[s][]{\{}{\}},
   moredelim=[s][\color{orange}]{f"}{"}
}

\newcommand\pythonstyle{\lstset{
    language=Python,
    frame=single,                                    
    columns=fullflexible,
    breaklines=true,
    captionpos=t,
    basicstyle=\fontsize{6pt}{5pt}\upshape\ttfamily,      
    numbers=left,                    
    numbersep=8pt,                   
    numberstyle=\tiny\color{gray}, 
    keywordstyle=\bfseries\color{blue!55!black},
    stringstyle=\color{orange},
    xleftmargin=14pt, 
    xrightmargin=6pt,
    commentstyle=\color{gray}\ttfamily,
    backgroundcolor=\color{white},
    morekeywords={self},              
    emph={MyClass,__init__,return,type, str, int,
               int, float, list, dict, bool,tuple},          
    emphstyle=\bfseries\color{green!50!black},    
    frame=tb,                         
    showstringspaces=false
}

}

\lstnewenvironment{python}[1][]
{
\pythonstyle
\lstset{#1}
}
{}

\newcommand\pythoninline[1]{{\pythonstyle\lstinline!#1!}}

\definecolor{addition}{rgb}{0, 0.6, 0} 
\definecolor{deletion}{rgb}{0.8, 0, 0} 
\definecolor{unchanged}{rgb}{0, 0, 0}  

\lstnewenvironment{pythondiff}[1][]
{
    \pythonstyle
    \lstset{
        showspaces=false,                          
        showtabs=false,                            
        columns=fixed,                             
        keepspaces=true,                           
        morecomment=[f][\color{deletion}]{-},       
        morecomment=[f][\color{addition}]{+},       
        morecomment=[f][\color{unchanged}]{ },      
        moredelim=[il][\color{deletion}\sout]{\-},  
        moredelim=[il][\color{addition}]{\+},       
        literate=                                  
        {-}{{\color{deletion}-}}1                  
        {+}{{\color{addition}+}}1                  
        {^}{{\ }}1 
        #1 
    }
}
{}

\newcommand\jacstyle{
    \lstdefinelanguage{jac}{
       keywords={sem, def, import, node, ignore, take, activity, exit, spawn, with,
               edge, walker, and, or, if, elif, else, for, with, by, while,
               continue, break, disengage, report, anchor, has, can, true, false, from,
               context, info, details, try, strict, length, test, digraph, subgraph, test, by, in, to,
               skip, assert, yield, class, obj, Method, can, glob, from, enum
           },
       sensitive=false, 
       morecomment=[l]{//}, 
       morecomment=[l]{\#}, 
       morecomment=[s]{/}{/}, 
       morestring=[b]", 
       morestring=[b]',
       morestring=[b]`
    }
    
    \lstset{
        language=jac,
        frame=single,
        columns=fullflexible,
        breaklines=true,
        captionpos=t,
        basicstyle=\fontsize{6}{5}\upshape\ttfamily,
        numbers=left,
        numbersep=8pt,
        numberstyle=\tiny\color{gray},
        keywordstyle=\bfseries\color{blue!55!black},
        stringstyle=\color{orange},
        xleftmargin=5pt,
        xrightmargin=10pt,
         commentstyle=\color{gray}\ttfamily,
        backgroundcolor=\color{white},
        morekeywords={self},
        emph={return,by,py,jac,entry,type, str, int,
               int, float, list, dict, bool},
        emphstyle=\bfseries\color{green!50!black},
        frame=tb,                         
        showstringspaces=false
    }

}

\newcommand\jacinline[1]{{\jacstyle\lstinline!#1!}}

\lstnewenvironment{jac}[1][]
    {
    \leavevmode
    \jacstyle
    \lstset{#1}
    }
{}

\lstdefinelanguage{output}{
    keywords={Output},
    basicstyle=\tiny\ttfamily, 
    breaklines=true,
    keywordstyle=\color{blue},
    stringstyle=\color{black},
    commentstyle=\color{gray},
    morecomment=[l]{//},
    columns=fullflexible,
    xleftmargin=15pt,
    xrightmargin=5pt,
    showstringspaces=false
}

\lstdefinelanguage{vision_output}{
    keywords={Output},
    basicstyle=\tiny\ttfamily, 
    breaklines=true,
    keywordstyle=\color{blue},
    stringstyle=\color{black},
    commentstyle=\color{gray},
    morecomment=[l]{//},
    columns=fullflexible,
    xleftmargin=5pt,
    xrightmargin=5pt,
    showstringspaces=false
}

\AtBeginDocument{%
  }

\begin{document}

\title{Prompt Less, Smile More: Semantic Engineering in Lieu of Prompt Engineering}

\author{Jayanaka L. Dantanarayana}
\orcid{0009-0000-4320-8280}
\affiliation{%
  \institution{University of Michigan}
  \city{Ann Arbor}
  \country{USA}
}
\email{jayanaka@umich.edu}

\author{Savini Kashmira}
\orcid{0009-0005-4911-7597}
\affiliation{%
  \institution{University of Michigan}
  \city{Ann Arbor}
  \country{USA}
}
\email{savinik@umich.edu}

\author{Xiaoyu Ma}
\orcid{0009-0003-5605-4729}
\affiliation{%
  \institution{University of Michigan}
  \city{Ann Arbor}
  \country{USA}
}
\email{xiaoyma@umich.edu}

\author{Krisztian Flautner}
\orcid{0009-0002-8347-1811}
\affiliation{%
  \institution{University of Michigan}
  \city{Ann Arbor}
  \country{USA}
}
\email{manowar@umich.edu}

\author{Lingjia Tang}
\orcid{0000-0002-5609-7775}
\affiliation{%
  \institution{University of Michigan}
  \city{Ann Arbor}
  \country{USA}
}
\email{lingjia@umich.edu}

\author{Jason Mars}
\orcid{0000-0002-7029-5292}
\affiliation{%
  \institution{University of Michigan}
  \city{Ann Arbor}
  \country{USA}
}
\email{profmars@umich.edu}

\renewcommand{\shortauthors}{Dantanarayana et al.}

\begin{abstract}
AI-integrated applications increasingly rely on large language models (LLMs), but their intended behavior is not always fully specified by code. Meaning Typed Programming (MTP) automates prompt construction from program structure, yet selection criteria, output requirements, and application policies can remain unstated. We introduce Semantic Engineering, a methodology for identifying such missing intent and selectively adding and refining model guidance. We realize this methodology through Semantic Context Annotations (SemTexts), a dedicated language mechanism independent of documentation and schema metadata. SemTexts associate natural-language instructions with individual program entities and use declaration scopes to determine where they apply. Our implementation extends the Jac compiler and MTP runtime to preserve these associations and select applicable guidance during automatic prompt construction. We evaluate Semantic Engineering on seven AI-integrated applications and two standard benchmarks across three models. Across the applications, it achieves a $5.46\times$ reduction in integration lines of code relative to Prompt Engineering while delivering equal or better task performance in most model--application settings, reaching up to $1.83\times$ the task performance of Prompt Engineering.
\end{abstract}

\maketitle

\section{Introduction}
\label{sec:intro}

\begin{figure*}
    \centering
    \includegraphics[width=\linewidth]{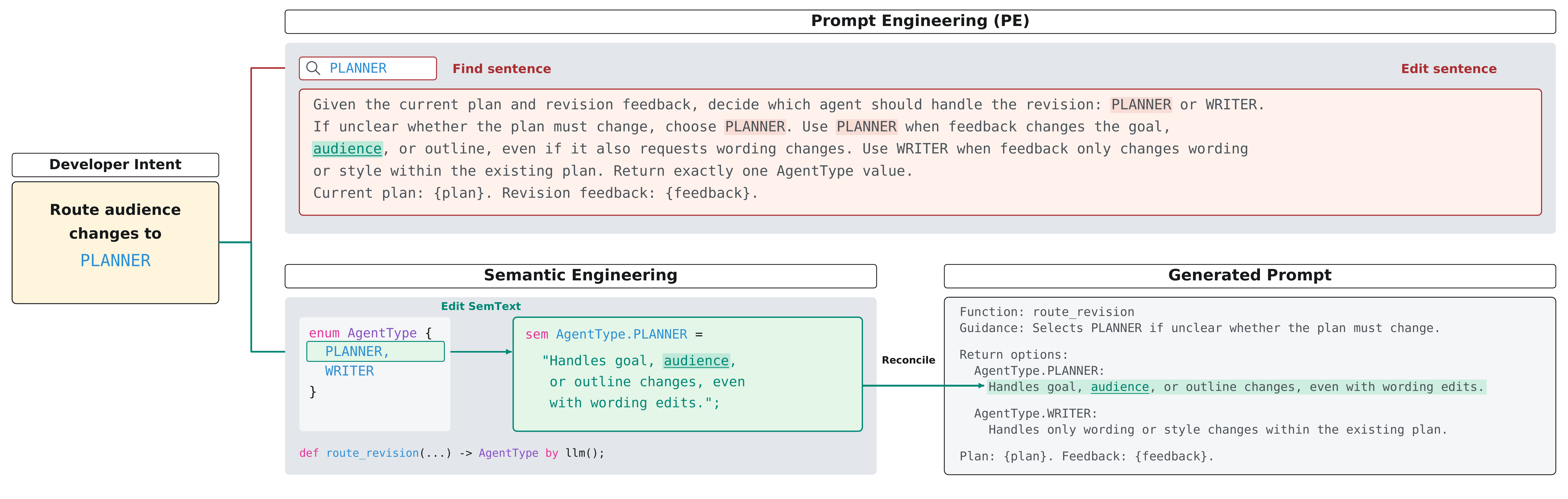}
    \caption{Implementing the same routing-policy change through Prompt Engineering and Semantic Engineering: routing audience changes to \texttt{PLANNER}. Prompt Engineering requires locating and editing the relevant sentence in a free-form prompt. Semantic Engineering targets the SemText bound to \texttt{AgentType.PLANNER}, which is automatically placed alongside the corresponding return option in the generated prompt.}
    \label{fig:intro}
    \Description{A developer's intent branches into two workflows: searching and editing a manual prompt, or editing the SemText associated with the PLANNER enum member. An arrow connects the edited SemText to its highlighted placement in the generated prompt.}
\end{figure*}

Large Language Models (LLMs) are increasingly integrated into software applications, including agentic systems~\cite{weber2024largelanguagemodelssoftware,scallop,lmql,sharma2025promptpexautomatictestgeneration}. These applications use LLMs for tasks such as extracting information, generating text and code, solving problems, planning, and making decisions. Prompt Engineering (PE) provides application developers with a direct way to guide how LLMs perform these tasks~\cite{sahoo2025systematicsurveypromptengineering}. When writing prompts manually, developers describe the task, provide relevant context, and specify the expected output. However, these prompts often repeat information already present in the code, such as field names, data types, and function interfaces. When developers change these definitions in the code, they must also update the corresponding descriptions in their prompts, adding complexity to application development and maintenance.

Frameworks such as DSPy~\cite{dspy} and NVIDIA Object-Oriented Agents (NOOA)~\cite{nooa2026furgale} reduce manual prompt construction through programming abstractions and library interfaces. Meaning Typed Programming (MTP)~\cite{mtp2025jayanaka} provides this support directly within the programming language. Through the \texttt{by llm} construct, developers express model functionality using existing code definitions, while the compiler and runtime automate prompt construction and response handling. The compiler determines which signatures, types, and identifiers contribute to each model request. MTP's evaluation demonstrates a reduction in the number of lines of code added or modified and the developer time required for LLM integration, while maintaining or improving task accuracy across the evaluated benchmarks~\cite{mtp2025jayanaka}.

However, the information available in code may not fully specify the intended model behavior. Function signatures and type definitions describe inputs and outputs, but may leave decision criteria, evidence requirements, and task policies unspecified. For example, code may define the available routing options without specifying the criteria for choosing among them. Such gaps can be addressed through targeted additions to the model's prompt that supply the missing intent while retaining the information already provided by the program. Developers therefore need a way to express this missing intent in code and include it in the model's prompt.

Existing frameworks already let developers selectively supply additional intent through comments, docstrings, and field descriptions included in model prompts. TypeChat uses comments in type definitions~\cite{typechat2023hejlsberg}, while DSPy supports field descriptions through Pydantic-based signatures~\cite{dspy,dspy2026fieldsource}. This text may also serve as code documentation or describe schema fields~\cite{pydantic2026schema}, so edits for those purposes can unintentionally change model behavior even when application requirements remain unchanged. Also, different functions may call the same LLM-guided function but require different instructions in the prompt generated for each call. If annotations are attached only to the shared function’s definition, developers need separate prompt configurations or context supplied to the model to specify which instructions apply to each call. These concerns motivate a dedicated model-guidance channel, independent of documentation and schema metadata, that lets developers declare instructions within the workflows where they apply. Using such a channel selectively requires identifying when code leaves developer intent unspecified and which code entity the additional guidance should be attached to, such as a function, type, or parameter.

To inform these choices, we studied AI-integrated applications in \S~\ref{sec:eval} to assess how selectively supplying intent missing from code affects model behavior. In the Service Center case study, four routing annotations increased end-to-end task success from 21.9\% to 85.6\%; adding two annotations for required reply facts raised success to 91.0\%. The six selected annotations outperformed full annotation and function-level docstrings, which achieved 64.7\% and 36.8\%, respectively. Adding an annotation that restated the reply composer's existing purpose provided only a further 0.2 percentage points. In Task Manager, supplying an approval requirement before sending email increased end-to-end task success from 70\% to 97\%, while leaving routing and reply composition unannotated. Together, these findings suggest that developers need support for selectively expressing intent missing from code at relevant code entities and placing that guidance alongside those entities in generated prompts.

\textbf{Our key insight is that a dedicated model-guidance channel, independent of documentation and schema metadata, should let developers selectively express intent left unspecified by code at the relevant program entities and preserve those associations in generated prompts.} The compiler binds each instruction to its target entity, and the runtime places applicable guidance alongside that entity's structural information. The channel also lets developers specify different additional instructions for calls to the same LLM-guided function, according to each caller’s requirements.

We introduce \textbf{Semantic Engineering}, an approach that selectively enriches code with explicit model guidance for behavior left unspecified by code. This guidance is placed precisely alongside the relevant code entities in generated prompts. We realize this approach through \textbf{Semantic Context Annotations (SemTexts)}, a language mechanism that extends MTP with optional model guidance bound to functions, parameters, types, fields, and enumeration members. Each annotation names its target entity and can be declared separately from that entity's definition; its declaration scope determines which calls can use it.

We implement SemTexts by extending the Jac compiler and MTP runtime~\cite{jac_cal,mars2025objectspatialprogramming,mars2025extendingdataspatialsemantics}. The compiler resolves annotation targets where possible using the program's scopes and symbol table, then preserves target references, guidance, and declaration scopes in an enriched \textbf{Meaning-Typed Intermediate Representation (MT-IR)}~\cite{mtp2025jayanaka}. The runtime completes any required target resolution and selects applicable guidance independently for each entity using nearest-scope precedence. Relevant bindings are checked for unresolved targets and duplicate declarations within the same scope before model invocation. The selected guidance is placed alongside the corresponding structural information in generated prompts. The pipeline governs how guidance is associated with program entities and delivered to model requests; interpreting the natural-language guidance remains the model's task.

We evaluate Semantic Engineering against base MTP, Prompt Engineering, and DSPy on seven AI-integrated applications and two standard benchmarks across three models. Across the applications, Semantic Engineering achieves a $5.46\times$ reduction in integration LOC relative to Prompt Engineering while delivering equal or better task performance in most model--application settings, reaching up to $1.83\times$ the task performance of Prompt Engineering. A separate coding-agent study, using token consumption as a proxy for cognitive effort, finds that Prompt Engineering consumes $1.55$--$1.86\times$ as many tokens as Semantic Engineering to reach the same development-test performance targets. In two application case studies, we further examine gradual semantic enrichment and compare selective enrichment with full annotation and docstring-based guidance.

This paper makes the following key contributions.
\begin{enumerate}
    \item We introduce \textbf{Semantic Engineering}, a methodology for identifying missing intent and selectively adding and refining model guidance, and realize it through \textbf{Semantic Context Annotations (SemTexts)}, a dedicated language mechanism for expressing guidance at program entities.

    \item We implement scoped binding, per-entity guidance selection, and validation in the Jac compiler and MTP runtime, preserving entity associations during prompt construction.

    \item We evaluate Semantic Engineering's implementation overhead, task performance, and coding-agent effort, and analyze selective enrichment through comparisons with full annotation and docstring-based guidance.
\end{enumerate}

\section{MOTIVATION}
\label{sec:motivating_example}

In this section, we present a motivating example to examine how developers express intent in AI-integrated programs. We first compare how developer intent is conveyed through prompts and code to identify what code leaves unspecified and where additional guidance is needed. We then examine existing ways to supply this guidance and their limitations. Finally, we illustrate language support for expressing missing intent through a dedicated channel and preserving its association with relevant code entities in generated prompts.

\subsection{Expressing Developer Intent through a Prompt}
\label{subsec:PE_impl}

\begin{figure}[tbp]
    \centering
    \begin{subfigure}[t]{0.48\linewidth}
        \centering
        \begin{python}
"""
Given the current plan and revision
feedback, select the agent that should
handle the revision: PLANNER or WRITER.

If feedback is ambiguous about whether
the plan must change, choose PLANNER.

Use PLANNER when feedback changes the
goal, audience, or outline, even if it
also requests wording changes.

Use WRITER when feedback only changes
wording or style within the existing plan.
"""
        \end{python}
        \caption{Illustrative manual prompt, showing only the instructions that express the intended behavior.}
        \label{fig:PE-prompt}
    \end{subfigure}\hfill
    \begin{subfigure}[t]{0.48\linewidth}
        \centering
        \begin{jac}
enum AgentType {
    PLANNER,
    WRITER
}

# Returns an identifier;
# the caller dispatches the agent.
def route_revision(
    plan: str,
    feedback: str
) -> AgentType by llm();
        \end{jac}
        \caption{MTP implementation of the routing function. Model configuration is omitted.}
        \label{fig:MTP-code}
    \end{subfigure}
    \caption{Developer intent expressed through a manual prompt and the corresponding MTP code. The code supplies the task interface and available options, while the selection and ambiguity rules require additional guidance.}
    \label{fig:motivation-prompt-code}
\end{figure}

To ground the discussion, we consider how an AI-Integrated content generation application selects an agent to handle a user's revision request. The application has a planning agent, which determines the goal, audience, and outline of the content, and a writing agent, which produces text based on that plan. After reviewing the generated text, a user may request changes. A routing function examines the current plan and the user's request to select which agent should act next: the planning agent if the plan requires revision, or the writing agent if the text can be revised within the existing plan.

Using Prompt Engineering, the developer can express the intended routing policy through the prompt in Figure~\ref{fig:PE-prompt}. For simplicity, we show only the task instructions, omitting input serialization, output-format instructions, and model invocation code. The prompt explains when to select each agent and directs the model to choose \texttt{PLANNER} when it is unclear whether the plan needs to change. For example, a request to ``Rewrite this for infrastructure engineers'' changes the audience of content originally planned for first-time users. The policy therefore selects \texttt{PLANNER}, even though the request asks for rewriting.

\subsection{What the Code Expresses and What It Leaves Unspecified}
\label{subsec:motivation-code-intent}

We next examine how developer intent expressed through code can guide the model's routing decision. We use MTP as the foundation for this discussion because it provides language-level constructs for model calls and generates prompts from function signatures and relevant type definitions~\cite{mtp2025jayanaka}. These capabilities make MTP a suitable foundation for exploring a dedicated guidance channel within the programming language, through which additional instructions can be associated with code entities and carried into generated prompts.

Figure~\ref{fig:MTP-code} shows an MTP implementation of \texttt{route\_revision}, the routing function that selects an agent to handle the user's revision request. The function takes the current plan and revision feedback and returns an \texttt{AgentType} value identifying the selected agent. The \texttt{by llm()} clause delegates this routing decision to an LLM, while MTP constructs the prompt using the program declarations and input values.

Comparing Figures~\ref{fig:PE-prompt} and~\ref{fig:MTP-code} shows which parts of the developer's intent are represented in code. The function and parameter names identify the routing task and its inputs, while the return type identifies the available agents. MTP includes this information in the generated prompt without requiring the developer to repeat it manually.

However, the selection criteria and ambiguity rule in the manual prompt are not expressed by these declarations. The code does not specify which changes require planning, whether planning takes precedence when feedback also requests rewriting, or how ambiguous feedback should be handled. Both \texttt{PLANNER} and \texttt{WRITER} satisfy the return type, but only \texttt{PLANNER} follows the stated policy when the request changes the audience. These rules express application-specific decisions that the function signature and enum definition alone do not determine.

The developer therefore needs a way to add the missing instructions while retaining MTP's ability to generate prompts from the information already expressed in code. This raises two questions: which parts of the intended behavior require guidance beyond what code already expresses, and at which code entities should that guidance be added?

\subsection{Expressing and Associating the Missing Guidance}
\label{subsec:motivation-guidance}

In the example shown in Figures~\ref{fig:PE-prompt} and~\ref{fig:MTP-code}, the additional guidance consists of the selection criteria and the rule for ambiguous feedback. The developer could place these rules in code comments, function docstrings, or field descriptions. An LLM integration could then include this text in its generated prompts. For example, a Pydantic field description can become part of a generated JSON Schema and reach the model when that schema is included in a request~\cite{pydantic2026schema}.

When supplying this guidance, the developer must also decide where each instruction belongs. The ambiguity rule applies to the routing function, while each selection criterion describes an individual routing option. Task instructions and field descriptions, such as those provided by DSPy~\cite{dspy,dspy2026fieldsource}, allow the developer to place the ambiguity rule in the signature's instructions and both selection criteria in the description of the \texttt{agent} output field. In this encoding, both selection criteria are attached to the \texttt{agent} output field; their associations with \texttt{PLANNER} and \texttt{WRITER} are expressed within the description's text. However, as the study summarized in the introduction shows, the placement of this guidance in the generated prompt can also affect model behavior. This motivates support for expressing guidance at the relevant code entities and preserving those associations during prompt generation.

Reusing existing text channels to express this guidance also raises a maintenance concern. When comments, docstrings, or field descriptions serve both ordinary documentation and model guidance, edits made for documentation can change the model's instructions. Suppose the \texttt{agent} field's description contains the selection criteria and is shared by API documentation and model requests. Shortening that description to ``The agent selected to handle revision feedback'' for documentation purposes also removes the criteria from the model request, even though the field's type and allowed values remain unchanged. The developer therefore needs to distinguish text maintained as model guidance from text maintained for documentation.

Different functions may also call the same LLM-guided function but require different instructions in the prompt generated for each call. If annotations are attached only to the shared function's definition, developers need separate prompt configurations or context supplied to the model to specify which instructions apply. This motivates allowing instructions to remain bound to the shared function while limiting their applicability to calls within a particular caller's scope.

In this example, developers need to supply the selection criteria and ambiguity rule while retaining the information already expressed in code. They also need to express where these instructions belong: the ambiguity rule applies to the routing function, while the selection criteria describe the \texttt{PLANNER} and \texttt{WRITER} enum members. These requirements motivate language support for adding missing guidance selectively at the relevant entities. A dedicated channel would identify these instructions as model guidance and allow them to be maintained independently of ordinary documentation. Prompt generation could then combine them with the information provided by code, preserving their associations with the relevant entities. This would let developers refine the additional guidance while retaining automatic prompt generation in language-level systems such as MTP.

\subsection{Semantic Engineering Vision}
\label{subsec:motivation-se-vision}

We envision language support that lets developers express intent not conveyed to the model by code, independently of documentation and schema metadata, and associate the additional instructions with the code entities they describe. In the MTP code in Figure~\ref{fig:MTP-code}, the function signature and enum definition already identify the task, its inputs, and the available agents. The developer adds the \texttt{sem} declarations shown in Figure~\ref{fig:motivation-semtexts} to this program to supply the missing routing policy. These declarations supplement the existing function and enum definitions.

\begin{figure}[tbp]
    \centering
    \begin{jac}
sem route_revision = "Selects PLANNER if unclear whether the plan must change.";
sem AgentType.PLANNER = "Handles goal, audience, or outline changes, even with wording edits.";
sem AgentType.WRITER = "Handles only wording or style changes within the existing plan.";
    \end{jac}
    \caption{Additional \texttt{sem} declarations for the MTP code in Figure~\ref{fig:MTP-code}. The declarations associate the ambiguity rule with the routing function and the selection criteria with their corresponding enum members.}
    \label{fig:motivation-semtexts}
\end{figure}

The declaration for \texttt{route\_revision} specifies how to handle ambiguity in the routing decision. The declarations for \texttt{PLANNER} and \texttt{WRITER} specify when each option applies, including the preference for \texttt{PLANNER} when feedback combines planning changes with wording edits. The developer therefore supplies the missing policy at two levels: the routing function and its individual return options. These declarations identify the additional text as model guidance, allowing it to be maintained independently of ordinary documentation.

These associations also determine how the guidance is presented to the model. The schematic prompt excerpts in Figure~\ref{fig:motivation-placement} contain identical guidance but place it differently. The first groups all guidance in the routing function's description. The second follows the declarations in Figure~\ref{fig:motivation-semtexts}, placing the ambiguity rule with the function and each selection criterion alongside its corresponding return option. Common input values and output-format instructions are omitted.

\begin{figure}[tbp]
    \centering
    \begin{subfigure}{\linewidth}
        \centering
        \begin{python}
"""
Function: route_revision
Guidance:
  Selects PLANNER if unclear whether the plan must change.
  AgentType.PLANNER:
    Handles goal, audience, or outline changes, even with wording edits.
  AgentType.WRITER:
    Handles only wording or style changes within the existing plan.

Return options: AgentType.PLANNER, AgentType.WRITER
"""
        \end{python}
        \caption{Guidance collected in the function description.}
        \label{fig:motivation-placement-grouped}
    \end{subfigure}

    \vspace{0.5em}

    \begin{subfigure}{\linewidth}
        \centering
        \begin{python}
"""
Function: route_revision
Guidance:
  Selects PLANNER if unclear whether the plan must change.

Return options:
  AgentType.PLANNER:
    Handles goal, audience, or outline changes, even with wording edits.
  AgentType.WRITER:
    Handles only wording or style changes within the existing plan.
"""
        \end{python}
        \caption{Guidance placed with the entities it describes.}
        \label{fig:motivation-placement-granular}
    \end{subfigure}
    \caption{Schematic prompt excerpts containing identical guidance. Their organization follows the association of guidance with the routing function or its individual return options.}
    \label{fig:motivation-placement}
\end{figure}

By expressing guidance at this granularity, developers can refine individual parts of the routing policy. A change to \texttt{sem AgentType.WRITER} revises the criterion for selecting that option, while a change to \texttt{sem route\_revision} revises the rule for ambiguous feedback. Prompt generation carries each revision alongside its associated entity. The same organization that determines where guidance appears therefore also gives developers specific places to review and refine how the program conveys intent to the model. Different functions may also call the same LLM-guided function but require different instructions in the generated prompt. For example, \texttt{standard\_workflow} and \texttt{editing\_workflow} may both call \texttt{route\_revision}, but direct ambiguous feedback to \texttt{PLANNER} and \texttt{WRITER}, respectively. Figure~6 shows how declarations within each caller's scope express these policies while retaining the shared enum-member guidance.

Building on this support, we envision Semantic Engineering as a development process for deciding what additional model guidance is needed, where it belongs, and how to refine it. Developers compare the application's routing criteria, output requirements, and policies with the information already available to the model, then express the missing requirements at the code entities they describe. Known gaps can be addressed before execution, while testing helps developers assess the resulting behavior and determine whether further guidance changes are needed. This process allows developers to enrich their programs selectively and gradually while retaining automatic prompt generation. \S~\ref{sec:sem_eng} develops this Semantic Engineering workflow.

\section{Semantic Engineering}
\label{sec:sem_eng}

We introduce Semantic Engineering as a methodology for selectively adding model guidance to AI-integrated applications. Algorithm~\ref{alg:semantic-enrichment} summarizes this methodology. Developers identify requirements not already conveyed to the model through code, inputs, or existing guidance (\S~\ref{subsec:3.1}), express those requirements as instructions with explicit target entities and declaration scopes (\S~\ref{subsec:3.2}), and refine the guidance through testing and diagnosis (\S~\ref{subsec:3.3}). The first two steps establish the initial guidance, while the third constitutes gradual semantic enrichment.

\subsection{Identifying Developer Intent Left Unspecified by Code}
\label{subsec:3.1}

To identify developer intent left unspecified by code, the developer compares the application's requirements for an LLM-guided operation with the information expressed through its identifiers, function signature, and types. The developer then checks whether requirements not expressed in the code are communicated through the supplied inputs or existing guidance. Any requirement still unstated needs to be expressed as natural-language guidance to help the model carry out the operation as intended. This comparison is the opening step of Algorithm~\ref{alg:semantic-enrichment}.

Consider a developer comparing the application's requirements with the code definitions in the revision-routing example (\S~\ref{sec:motivating_example}). The requirements specify that feedback requiring changes to the plan should be handled by \texttt{PLANNER}, while feedback requesting only wording changes should be handled by \texttt{WRITER}. The signature of \texttt{route\_revision} and the definition of \texttt{AgentType} identify the inputs and available agents, but do not express these rules for choosing between them. If the supplied plan, feedback, and existing guidance also leave these rules unstated, the developer identifies them as missing intent. This gap can be identified before execution, without first observing an incorrect choice.

In this routing example, the return type identifies the available agents but does not state the criteria for selecting one. Developers can look for similar omissions in other LLM-guided decisions, including action selection, approval, and termination. The relevant requirements specify when an action should be selected, what makes an output acceptable, or when execution should stop. For example, a Boolean return type for an approval operation represents acceptance or rejection but does not specify the conditions for acceptance. Such application-specific criteria are useful starting points for identifying missing intent because function signatures and types do not necessarily express them. The developer can check these requirements against the code, supplied inputs, and existing guidance to determine which criteria still need to be expressed through semantic enrichment.

\subsection{Adding Missing Guidance to Program Entities}
\label{subsec:3.2}

For each requirement identified in the previous step, the developer expresses the missing intent as natural-language guidance that supplements the information already available to the model. This guidance can specify a decision criterion, clarify the interpretation of a value, or state a requirement on an output. In the routing example, the guidance can state that \texttt{PLANNER} should be selected when feedback requires changes to the plan.

The developer adds this guidance to the program by associating it with the entity whose meaning or behavior it clarifies. In the routing example, the developer associates the rule for resolving ambiguous feedback with \texttt{route\_revision}, because the rule governs how the function chooses between agents. The criteria for selecting \texttt{PLANNER} and \texttt{WRITER} are associated with their respective enum members, because they describe when each option applies. Similarly, an instruction about a parameter's interpretation is associated with that parameter, and a requirement on an output field is associated with that field. These associations guide the placement of each instruction alongside the corresponding entity in generated prompts.

When different workflows require different instructions for a shared function, the developer associates each instruction with that function and declares it within the lexical scope containing the calls that should use it. For example, two workflow functions may both call \texttt{route\_revision} but require different rules for ambiguous feedback. The developer declares each rule within the corresponding workflow function's scope, so that both workflows can reuse the same routing function with the guidance intended for their own calls. These steps correspond to the first loop in Algorithm~\ref{alg:semantic-enrichment}.

\subsection{Gradual Semantic Enrichment}
\label{subsec:3.3}

After initial enrichment, some application requirements may still not be conveyed to the model, and some of the added instructions may be unclear. Gradual semantic enrichment therefore adds a cycle of testing, diagnosis, and revision to check the resulting behavior and determine whether additional or revised guidance is needed. This cycle corresponds to the refinement loop in Algorithm~\ref{alg:semantic-enrichment}.

The developer begins this cycle by running representative tests and comparing the LLM-guided decisions and application outcomes with the requirements. When behavior differs from what is expected, the developer checks for omitted requirements, unclear or conflicting instructions, and guidance attached to the wrong entity or declared in the wrong scope. If the diagnosis supports a guidance change, the developer revises the relevant instruction and tests again. Failures caused by other parts of the implementation or system require changes to those parts.

Identifying missing requirements and diagnosing failures remain developer judgments; the compiler and runtime provide the binding, guidance-selection, and prompt-construction mechanisms described in Section~4. The developer stops revising guidance when the acceptance criteria are met or the diagnosis does not justify another guidance change. Remaining issues are recorded for further investigation. As development continues, requirements may change and some guidance may become redundant. The developer can update or remove individual instructions and repeat the testing cycle, without requiring every entity to be annotated.

\begin{algorithm}[h]
\footnotesize
\caption{Semantic Engineering}
\label{alg:semantic-enrichment}

\KwIn{Application requirements $R$, program $P$, test cases $\mathcal{T}$}
\KwOut{Updated program $P$, test results $E$, unresolved observed issues $U$}

\tcp{Initial enrichment from the specification}
Compare $R$ with the code information, inputs, and existing guidance supplied to the relevant LLM-guided operations\;

\ForEach{requirement identified by the developer as needing additional guidance}{
    Express the requirement as concise natural-language guidance $q$\;
    Select the program entity $e$ whose meaning or behavior $q$ clarifies\;
    Select the declaration scope $s$ where the guidance should apply\;
    Add guidance $q$ for entity $e$ in scope $s$ to $P$\;
}

\tcp{Refinement through testing}
\While{true}{
    Run $\mathcal{T}$ on $P$ and record results $E$\;
    Compare the decisions and application outcomes with $R$, recording unresolved observed issues in $U$\;

    \If{the chosen acceptance criteria are met}{
        \Return{$(P, E, U)$}\;
    }

    Diagnose the differences between observed and intended behavior\;

    \eIf{the diagnosis supports a specific change to the guidance}{
        Add, revise, or remove guidance in $P$, adjusting its text, associated entities, or scopes as needed\;
    }{
        \Return{$(P, E, U)$}\;
    }
}

\end{algorithm}

Supporting this workflow while retaining automatic prompt construction leads to the following design requirements for the language mechanism and its implementation:

\begin{enumerate}
    \item \textbf{Dedicated model guidance.} Provide a dedicated way to express natural-language instructions for the model, allowing developers to maintain guidance independently of ordinary documentation.

    \item \textbf{Explicit entity associations.} Associate each instruction with the program entity it describes, including functions, parameters, types, fields, and individual enum members.

    \item \textbf{Scoped guidance.} Allow guidance to be declared in the scope containing the calls that should use it, independently of where the target entity is defined. This allows different workflows to supply their own guidance for a shared entity.

    \item \textbf{Preservation during prompt construction.} Preserve each instruction's entity association and scope through compilation and execution, so that applicable guidance can be selected and placed alongside the corresponding code information in automatically constructed prompts.
\end{enumerate}

These requirements motivate Semantic Context Annotations (SemTexts), our language mechanism for realizing Semantic Engineering. Section~4 presents the SemText syntax and binding rules and explains how the compiler and runtime implement these requirements.

\section{SemTexts: Design and Implementation}
\label{sec:gse}

SemTexts realize the workflow in \S~\ref{sec:sem_eng} by extending MTP with declarations for application requirements left unspecified by code. Each declaration associates natural-language guidance with a program entity, while its declaration scope determines where the guidance applies. The compiler and runtime use these associations and scopes to place applicable guidance alongside the corresponding entities in automatically constructed prompts.

\begin{figure}[t]
    \centering
    \includegraphics[width=\linewidth]{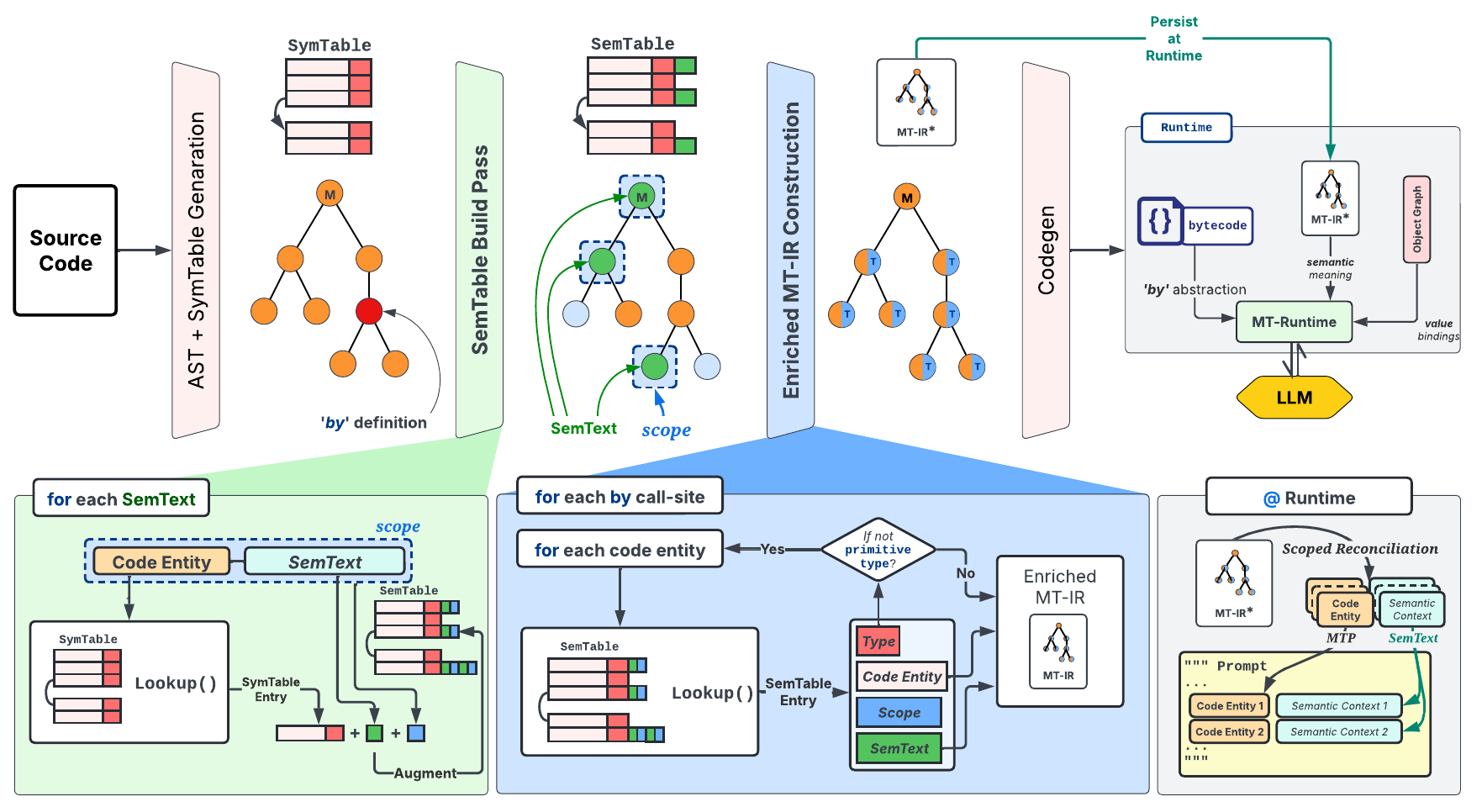}
   \caption{Compiler and runtime support for SemTexts. The SemTable records guidance together with target references and declaration scopes. Enriched MT-IR retains the code structure, candidate bindings, and scope metadata needed for guidance selection. At runtime, applicable guidance is placed alongside its corresponding entities and combined with current input values.}
    \label{fig:compilerandrun}
\end{figure}

Figure~\ref{fig:compilerandrun} summarizes the compiler and runtime pipeline for incorporating SemText guidance into generated prompts. The compiler records each SemText's target reference, guidance string, and declaration scope in a SemTable, retaining partially resolved references for later resolution. For each LLM-guided call site, it combines the relevant code structure with these bindings in an enriched Meaning-Typed Intermediate Representation (MT-IR). This representation is preserved through code generation, making the entity associations and scope information available at runtime. The runtime then selects the guidance applicable to the call and combines it with the code structure and current input values to construct the model's prompt.

\subsection{SemText Declarations: Targets and Scope}
\label{subsec:semtex_syntax}

A SemText declaration associates natural-language guidance with a program entity using the \texttt{sem} keyword. Its schematic syntax is:
\begin{equation}
    S \rightarrow \texttt{sem}\ T\ \texttt{=}\ Q\ \texttt{;}
    \label{eq:semtext-syntax}
\end{equation}

\begin{figure}[t]
\begin{jac}
sem route_revision = "Selects PLANNER if unclear whether the plan must change.";

def standard_workflow(plan: str, feedback: str) -> AgentType {
    return route_revision(plan, feedback);
}

def editing_workflow(plan: str, feedback: str) -> AgentType {
    sem route_revision = "If plan changes are unclear, assume wording-only and use WRITER.";
    return route_revision(plan, feedback);
}
\end{jac}
\caption{Guidance for a shared routing function declared at module and workflow scopes. The routing function and return type are defined in Figure~\ref{fig:MTP-code}; the shared enum-member guidance is shown in Figure~\ref{fig:motivation-semtexts}.}
\label{fig:scoped-routing}
\end{figure}

Here, $T$ identifies the target program entity, and $Q$ is a string literal containing the guidance. The target can be a named declaration defined in or imported into the program, such as a function, method, or type, or a named component of such a declaration, such as a parameter, field, or enum member. For example, $T$ can name the function \texttt{route\_revision}, as shown in Figure~\ref{fig:scoped-routing}, to associate guidance with that function.

Where the developer writes this \texttt{sem} declaration determines which calls can use the guidance attached to $T$. For example, to supply guidance specifically for the call to \texttt{route\_revision} inside \texttt{editing\_workflow}, the developer writes the declaration inside that workflow's function body, as shown in Figure~\ref{fig:scoped-routing}. The declaration names \texttt{route\_revision} as its target, while its lexical scope makes the guidance local to \texttt{editing\_workflow}. The declaration at module scope supplies the default guidance used by \texttt{standard\_workflow}. This allows guidance for a shared function to be declared within the workflows that use it, separately from the function's definition.

To let the runtime select the guidance applicable to each call, the compiler records each declaration's target reference, guidance string, and declaration scope in a \emph{SemTable}, as shown in the green panel at the lower left of Figure~\ref{fig:compilerandrun}. Any unresolved part of a target reference is retained for resolution when the target is needed at an LLM-guided call site. \S~\ref{subsec:compilation} explains how these bindings are preserved during compilation, while \S~\ref{subsec:runtime} explains how the runtime uses them to select applicable guidance and place it in the generated prompt.

\subsection{Compiling SemTexts into Enriched MT-IR}
\label{subsec:compilation}

\paragraph{Recording SemText bindings.}
After parsing and symbol analysis, the compiler uses the program's symbol table and available import information to resolve each SemText target as far as possible. It records the resulting target reference together with the guidance string and the scope containing the \texttt{sem} declaration. The green panel at the lower left of Figure~\ref{fig:compilerandrun} shows this recording step. Each resulting binding has the form:
\begin{equation}
    b = \langle s,r,q\rangle,
    \label{eq:semantic-binding}
\end{equation}
where $s$ is the SemText declaration scope, $r$ is the target reference, and $q$ is the guidance string. A fully resolved reference identifies a program entity $e$. A partially resolved reference preserves the information already known about the target together with the names that still require resolution. The SemTable stores the binding as $\Sigma(s,r)=q$, using the declaration scope and target reference to identify the entry.

The following rules govern how SemText bindings are recorded and resolved:
\begin{enumerate}
    \item \textbf{Supported targets.} A target must name a declaration defined in or imported into the program, or a named component of such a declaration, as described in \S~\ref{subsec:semtex_syntax}. Expressions that access individual runtime values, such as \texttt{items[1]}, and attributes synthesized through dynamic lookup are not supported targets. Generic type applications are also excluded as annotation targets; their constituent types are followed during structural construction.

    \item \textbf{Deferred target resolution.} An imported type may be identifiable before information about its fields is available. A SemText targeting one of those fields is retained without requiring complete resolution during compilation. When the type or member is needed at an LLM-guided call site, the structural information available for that call is used to complete the relevant target resolution. A declaration whose target is not needed by a call remains recorded without contributing guidance to that call. Under the lexical-scope rules described in \S~\ref{subsec:runtime}, SemTexts declared in an imported type's defining module do not apply to a call unless their declaration scope lies on that call's recorded lexical scope chain.

    \item \textbf{Uniqueness within a scope.} Duplicate SemText declarations for the same entity in the same scope are rejected, with the check deferred when target resolution is incomplete. Declarations in different scopes remain separate bindings.
\end{enumerate}

For example, the module-level declaration in Figure~\ref{fig:scoped-routing} and the declaration inside \texttt{editing\_}-\texttt{workflow} both target \texttt{route\_revision}. As they occur in different scopes, the compiler records separate SemTable entries, each retaining the corresponding target reference, declaration scope, and guidance string.

\paragraph{Identifying entities for enrichment.}
With the SemText bindings recorded, the compiler next uses the called operation's structure to identify which bindings may contribute guidance. This step reuses MTP's existing structural representation, denoted $\operatorname{MTIR}(f)$ for an operation $f$, which contains its interface and relevant type structure~\cite{mtp2025jayanaka}. The entity and type references in this representation connect the operation's structure to the bindings retained in the SemTable.

The compiler follows available type references, including generic type arguments, to expose relevant types and their named components. This traversal proceeds whether or not the containing type has a SemText declaration. When an imported type's structure is unavailable, its reference is retained so that expansion can continue when the required information becomes available for a call.

For \texttt{route\_revision}, expanding the \texttt{AgentType} return type exposes \texttt{PLANNER} and \texttt{WRITER} as separate entities. The subsequent enrichment step can therefore collect candidate guidance for each member even when \texttt{AgentType} itself has no SemText declaration. Following the type structure thus makes guidance on individual components available alongside the components they describe.

\paragraph{Constructing enriched MT-IR}
The compiler uses the program entities identified during structural traversal and the type references retained for later expansion to collect candidate SemText bindings from the SemTable. For a source call site $c$ invoking operation $f$, let $\mathcal{E}_f$ denote the entities currently represented in $\operatorname{MTIR}(f)$. The collected bindings are stored in a map $B_c$, where each entry $B_c(s,r)=q$ retains the declaration scope, target reference, and guidance string.

A binding is retained if its target reference identifies an entity in $\mathcal{E}_f$, or if incomplete target resolution or type information prevents ruling out its relevance to the operation. Bindings known to be unrelated are omitted. Collection considers all declaration scopes, so $B_c$ may contain several candidate SemTexts for the same entity. These alternatives remain available for runtime selection.

To support that selection, the compiler records the lexical scope containing the call, $s_c=\operatorname{scope}(c)$, and its enclosing-scope relationships, $\rho_c$. Here, $\rho_c(s)$ gives the lexical parent of scope $s$, with $\rho_c(s)=\bot$ when $s$ has no parent. The enriched representation is:
\begin{equation}
    \operatorname{MTIR}^{*}(c)
    =
    \langle
        \operatorname{MTIR}(f),
        B_c,
        s_c,
        \rho_c
    \rangle.
    \label{eq:enriched-mtir}
\end{equation}

Algorithm~\ref{alg:construct-enriched-mtir} summarizes this construction, corresponding to the blue panel in the lower middle of Figure~\ref{fig:compilerandrun}. It first invokes MTP's existing structural construction, then collects candidate bindings and records scope metadata. The predicate $\operatorname{MayContribute}(r,M,\mathcal{S})$ implements the retention rule above using the available symbol and type information $\mathcal{S}$: it returns true for a target represented in $M$ or one whose relevance remains undetermined.

\begin{algorithm}[t]
\footnotesize
\caption{Constructing Enriched MT-IR}
\label{alg:construct-enriched-mtir}

\KwIn{Source call site $c$ invoking $f$, symbol and type information $\mathcal{S}$, SemTable $\Sigma$}
\KwOut{Enriched MT-IR $M_c^{*}$}

$M \leftarrow \operatorname{BuildStructuralMTIR}(f,\mathcal{S})$\;

$B_c \leftarrow \emptyset$\;
\ForEach{SemTable entry $\Sigma(s,r)=q$}{
    \If{$\operatorname{MayContribute}(r,M,\mathcal{S})$}{
        $B_c(s,r) \leftarrow q$\;
    }
}

$s_c \leftarrow \operatorname{scope}(c)$\;
$\rho_c \leftarrow \operatorname{EnclosingScopeMetadata}(s_c,\mathcal{S})$\;
$M_c^{*} \leftarrow \langle M,B_c,s_c,\rho_c\rangle$\;
\Return{$M_c^{*}$}\;

\end{algorithm}

Code generation preserves the enriched MT-IR for runtime lookup. At invocation, the runtime identifies the calling function or module from the execution context and locates the corresponding compiled scope metadata, without traversing the source AST. \S~\ref{subsec:runtime} explains how it completes any required target resolution and selects guidance from the retained candidates.

\subsection{Selecting SemTexts and Constructing Prompts at Runtime}
\label{subsec:runtime}

Using the enriched MT-IR and scope metadata located for the current invocation, the runtime resolves the candidate bindings needed by the call. As illustrated in the lower-right runtime panel of Figure~\ref{fig:compilerandrun}, it then selects the applicable SemTexts and constructs the model's prompt by placing their text alongside the corresponding structural entries and incorporating the current input values.

\paragraph{Resolving candidate bindings.}
The compiler-retained bindings have the form $B_c(s,r)=q$, where a target reference $r$ may still be partially resolved. The runtime first considers bindings whose declaration scopes lie on the call's recorded lexical scope chain. It continues MTP's structural traversal where necessary, using the declared type and member metadata of loaded imported definitions to expand retained type references. Let $\mathcal{E}_c$ denote the entities represented in the request after this expansion.

The runtime resolves candidate references that name entities in this structure or named components of types reached by the traversal. It records the resulting bindings in a map $\widehat{B}_c(s,e)=q$, preserving each SemText's declaration scope and guidance string.

As target identities become known, the runtime performs any outstanding uniqueness checks and rejects duplicate declarations for the same entity in the same scope. If a candidate reference required by the request's structure remains unresolved, the runtime reports a resolution error before invoking the model. Retained candidates outside the call's lexical scope chain or unrelated to its structure do not cause resolution errors for that call.

\paragraph{Selecting applicable SemTexts.}
Algorithm~\ref{alg:resolve-semtext} defines $\operatorname{ResolveSem}$, which selects a SemText for one entity from the resolved candidate bindings. The algorithm starts at the call's lexical scope $s_c$ and follows the recorded parent relationships $\rho_c$. It returns the first matching SemText, implementing nearest-scope precedence. If no binding is found along the enclosing-scope chain, it returns $\bot$.

\begin{algorithm}[t]
\footnotesize
\caption{Selecting a SemText for an Entity (\textsc{ResolveSem})}
\label{alg:resolve-semtext}

\KwIn{Entity $e$, call-site scope $s_c$, resolved candidate bindings $\widehat{B}_c$, enclosing-scope metadata $\rho_c$}
\KwOut{Selected SemText string $q$, or $\bot$}

$s \leftarrow s_c$\;
\While{$s \neq \bot$}{
    \If{$(s,e)\in\operatorname{dom}(\widehat{B}_c)$}{
        \Return{$\widehat{B}_c(s,e)$}\;
    }
    $s \leftarrow \rho_c(s)$\;
}
\Return{$\bot$}\;

\end{algorithm}

The runtime applies this procedure independently to each entity represented in the request:
\begin{equation}
    \Gamma_c
    =
    \left\{
        e \mapsto
        \operatorname{ResolveSem}
        (e,s_c,\widehat{B}_c,\rho_c)
        \;\middle|\;
        e\in\mathcal{E}_c
    \right\}.
    \label{eq:call-guidance}
\end{equation}
A result of $\bot$ leaves the entity's structural description without an additional SemText. This denotes the absence of an applicable declaration, rather than a failure to resolve its target.

\paragraph{Constructing the prompt.}
Using the selection results in $\Gamma_c$, the runtime constructs the prompt from the operation's structural entries, their selected SemTexts, and the current input values. A function's SemText accompanies the operation description; a parameter's SemText accompanies the corresponding input description; and a SemText for a type, field, or enum member accompanies that entity's structural entry. Existing MTP machinery supplies the surrounding request structure and value serialization.

The binding and selection rules do not establish precedence between instructions attached to different entities or resolve contradictions between them; interpreting the selected guidance remains the model's task.

For the calls in Figure~\ref{fig:scoped-routing}, the runtime selects the module-level SemText for \texttt{route\_revision} when called from \texttt{standard\_workflow}, and the local SemText for the same function when called from \texttt{editing\_workflow}. The local declaration changes the instruction associated with \texttt{route\_revision} in the editing workflow's prompt. Because \texttt{route\_revision} returns \texttt{AgentType}, the runtime includes \texttt{AgentType.PLANNER} and \texttt{AgentType.WRITER} as return options, together with their SemTexts from Figure~\ref{fig:motivation-semtexts} in \S~\ref{sec:motivating_example}. These SemTexts describe when to select each agent and appear in both workflows' prompts. The local declaration in \texttt{editing\_workflow} changes the function's instruction for handling ambiguous feedback; it does not change the descriptions of the return options.

\subsection{SemText Implementation}
\label{subsec:semtexts-implementation}

The implementation extends the Jac compiler and MTP runtime~\cite{mtp2025jayanaka,jac_cal}, adding support for SemText declarations, scope-aware bindings, and their incorporation into generated prompts.

\paragraph{Compiler integration.}
The Jac parser is extended to retain \texttt{sem} declarations in the AST. A SemTable pass builds on Jac's existing symbol and import information, while MT-IR construction adds candidate bindings and scope metadata to MTP's structural representation. Code generation preserves this enriched representation for runtime access. These extensions implement the compilation procedure described in \S~\ref{subsec:compilation}.

\paragraph{Runtime integration.}
The MTP runtime is extended to access the preserved bindings and scope metadata and apply the resolution and selection procedures in \S~\ref{subsec:runtime}. Its prompt renderer incorporates selected SemTexts alongside the corresponding structural entries. Existing MTP components continue to handle value serialization, model invocation, and response processing.

\section{Evaluation}
\label{sec:eval}

We evaluate Semantic Engineering (SemEng) to determine whether adding \textit{SemTexts} to MTP improves application performance while preserving concise implementations. We address the following research questions:

\begin{enumerate}[label=\textbf{RQ\arabic*:}]
    \item \label{RQ1} How does Semantic Engineering affect implementation overhead in AI-integrated applications?
    \item \label{RQ2} How does Semantic Engineering compare with baseline approaches in terms of task performance?
    \item \label{RQ3} How does Semantic Engineering compare with Prompt Engineering in the cognitive effort required to express missing application intent and reach a common performance target?
    \item \label{RQ4} Where does code leave developer intent unspecified, and how does Semantic Engineering address these gaps to improve application behavior?
\end{enumerate}

\begin{table*}[t]
\centering
\footnotesize
\caption{Benchmark applications, datasets, evaluation methods, and applicable AI-integrated capabilities. C1--C6 are defined below the table.}
\label{tab:benchmarks}
\renewcommand{\arraystretch}{1.15}
\setlength{\tabcolsep}{4pt}
\begin{tabularx}{\textwidth}{
    |>{\raggedright\arraybackslash}p{1.5cm}
    |>{\raggedright\arraybackslash}X
    |>{\raggedright\arraybackslash}p{3.0cm}
    |>{\raggedright\arraybackslash}p{2.0cm}|
}
\hline
\textbf{Benchmark Application}
& \textbf{Task Description and Dataset}
& \textbf{Evaluation Method}
& \textbf{Capabilities} \\
\hline

Memory Retrieval
& Retrieve relevant memories from a user's personal memory database to answer a query. \textit{Dataset: 20 user profiles with 10 memories and five questions per user; 100 question instances, including 95 unique instances.}
& Macro-averaged F1, averaged over three runs.
& C3, C4 \\
\hline

Image Extraction
& Extract structured information and describe visual features from images. \textit{Dataset: A fixed manifest of 300 images sampled from LAION-400M~\cite{laion-image}.}
& Hybrid score combining semantic similarity through cosine similarity and lexical matching through BM25.
& C1, C4 \\
\hline

Task Manager
& Manage task creation and completion, task summarization, and email drafting and sending. \textit{Dataset: 220 synthetic queries comprising 100 task-handling, 20 task-summary, and 100 email-handling requests.}
& LLM-judged success rate using a fixed rubric that checks routing, request-specific validity, and applicable date calculations.
& C2, C3, C4, C5 \\
\hline

Content Creator
& Perform writing tasks through a multi-agent workflow with planning, writing, and reviewing stages. \textit{Dataset: 200 InstructEval prompts~\cite{chia2023instructeval} across four writing categories: informative, professional, argumentative, and creative.}
& LLM-judged success rate. A response is successful when it satisfies at least 11 of 15 evaluation criteria.
& C3, C4, C5, C6 \\
\hline

Aider Genius
& Resolve software issues through a planner that guides the Aider code editor in modifying an existing repository. \textit{Dataset: 60 SWE-bench Lite instances~\cite{jimenez2024swebench}.}
& Percentage of instances resolved under the SWE-bench test-based evaluation.
& C2, C3, C5, C6 \\
\hline

Service Center
& Resolve customer-service tickets using a router, seven specialist agents with distinct toolsets, and a reply merger. Specialists handle refunds, cancellations, status checks, credentials, email, rescheduling, and escalation. \textit{Dataset: A frozen set of 240 synthetic ticket episodes executed against a mock backend.}
& End-to-end pass rate, determined by comparing the resulting backend state with the expected canonical state, following $\tau$-bench~\cite{yao2024tau}.
& C2, C5, C6 \\
\hline

Fact Verification
& Verify claims in news articles through routing, seeding, relaying, and decision stages. Agents query a claim database and assign true, false, or unverifiable verdicts. \textit{Dataset: 20 articles with approximately 20 reference claims per article and a PostgreSQL-backed claim database. Claims and databases from ClaimDB~\cite{theologitis-etal-2026-claimdb}, built on BIRD~\cite{li2023bird}.}
& End-to-end verdict accuracy over all reference claims.
& C2, C4, C5, C6 \\
\hline

\end{tabularx}

\vspace{3pt}
\begin{minipage}{\textwidth}
\footnotesize
\textbf{C1:} Multimodal Perception \quad \textbf{C2:} Tool Use \& Integration \quad \textbf{C3:} Memory \& Context Management \quad \textbf{C4:} Information Retrieval \quad \textbf{C5:} Multi-Agent Coordination \quad \textbf{C6:} Planning \& Reasoning
\end{minipage}
\end{table*}

\paragraph{\textbf{Experimental Setup}}
We implement SemEng through \textit{SemTexts} in the Jac programming environment~\cite{jac_cal, mars2025objectspatialprogramming, mars2025extendingdataspatialsemantics}, extending the MTP compiler and runtime to incorporate entity-associated annotations into generated prompts. Our application experiments use Claude Opus 5.5, DeepSeek V4.1 Flash, and Gemma~3~27B. For Task Manager and Content Creator, we use GPT-4.1-mini as an LLM judge under fixed, application-specific evaluation rubrics.

\paragraph{\textbf{Baselines}}
We compare four integration approaches. 

\begin{enumerate}
    \item \textbf{Prompt Engineering (PE).} We use traditional prompt engineering, with manually written prompts and explicit model calls to integrate LLM functionality into the application.

    \item \textbf{Base MTP}~\cite{mtp2025jayanaka}. We use standard MTP, which relies on the semantics conveyed by the code to generate prompts.

    \item \textbf{DSPy}~\cite{dspy}. The same application requirements captured by SemTexts are expressed through signature descriptions and input/output field annotations, which DSPy incorporates into model requests.

    \item \textbf{Semantic Engineering (MTP + \textit{SemTexts}).} We add SemTexts to relevant program entities in Base MTP to express application requirements not explicit in the code. MTP incorporates these SemTexts into generated prompts.
\end{enumerate}

For each application, we ensure that the prompts used by PE, DSPy, and SemEng provide the model with the same application requirements and task information. To achieve this, we review the final prompts produced by these three approaches and sent to the model, both manually and with LLM assistance. We refine the implementations to include any missing requirements and verify the revised prompts through both forms of review.

\paragraph{\textbf{Benchmark Applications}}
We evaluate SemEng on GSM8K~\cite{cobbe2021gsm8k} for mathematical reasoning and HotpotQA~\cite{yang2018hotpotqa} for multi-hop question answering. These benchmarks provide established tasks for evaluating performance, but do not cover the full range of capabilities we examine in AI-integrated applications: \textbf{C1:} multimodal perception, \textbf{C2:} tool use and integration, \textbf{C3:} memory and context management, \textbf{C4:} information retrieval, \textbf{C5:} multi-agent coordination, and \textbf{C6:} planning and reasoning. We therefore construct a suite of seven application benchmarks, each exercising one or more of these capabilities and collectively covering all six. This suite allows us to examine how SemEng affects application behavior when model-guided operations interact with tools, application state, or other agents. Table~\ref{tab:benchmarks} summarizes the applications, datasets, evaluation methods, and capability coverage.

\subsection{RQ1: Implementation Overhead of Semantic Engineering}

\begin{wraptable}{r}{0.68\textwidth}
\vspace{-0.4cm}
\centering
\caption{Implementation overhead measured as LOC added or modified for LLM integration. Subscripts report the ratio of PE's LOC to the corresponding implementation's LOC. The final row reports the arithmetic mean of these per-application ratios.}
\label{tab:loc}

\footnotesize
\renewcommand{\arraystretch}{1.25}
\setlength{\tabcolsep}{2.5pt}

\begin{adjustbox}{max width=\linewidth}
\begin{tabular}{@{}lcccc@{}}
\toprule

\multirow{2}{*}{\textbf{Benchmark}}
& \textbf{PE}
& \textbf{Base MTP}
& \textbf{MTP + SemTexts}
& \textbf{DSPy} \\

\cmidrule(lr){2-2}
\cmidrule(lr){3-3}
\cmidrule(lr){4-4}
\cmidrule(lr){5-5}

& \textbf{LOC}
& \textbf{LOC$_{\downarrow\times}$}
& \textbf{LOC$_{\downarrow\times}$}
& \textbf{LOC$_{\downarrow\times}$} \\

\midrule

Memory Retrieval
& 65
& $4_{\textcolor{green!65!black}{\downarrow\!\times16.25}}$
& $6_{\textcolor{green!65!black}{\downarrow\!\times10.83}}$
& $22_{\textcolor{green!65!black}{\downarrow\!\times2.95}}$ \\

Image Extraction
& 126
& $11_{\textcolor{green!65!black}{\downarrow\!\times11.45}}$
& $18_{\textcolor{green!65!black}{\downarrow\!\times7.00}}$
& $34_{\textcolor{green!65!black}{\downarrow\!\times3.71}}$ \\

Task Manager
& 145
& $16_{\textcolor{green!65!black}{\downarrow\!\times9.06}}$
& $28_{\textcolor{green!65!black}{\downarrow\!\times5.18}}$
& $107_{\textcolor{green!65!black}{\downarrow\!\times1.36}}$ \\

Content Creator
& 130
& $21_{\textcolor{green!65!black}{\downarrow\!\times6.19}}$
& $43_{\textcolor{green!65!black}{\downarrow\!\times3.02}}$
& $115_{\textcolor{green!65!black}{\downarrow\!\times1.13}}$ \\

Aider Genius
& 189
& $25_{\textcolor{green!65!black}{\downarrow\!\times7.56}}$
& $39_{\textcolor{green!65!black}{\downarrow\!\times4.85}}$
& $233_{\times0.81}$ \\

Service Center
& 276
& $29_{\textcolor{green!65!black}{\downarrow\!\times9.52}}$
& $93_{\textcolor{green!65!black}{\downarrow\!\times2.97}}$
& $236_{\textcolor{green!65!black}{\downarrow\!\times1.17}}$ \\

Fact Verification
& 318
& $20_{\textcolor{green!65!black}{\downarrow\!\times15.90}}$
& $73_{\textcolor{green!65!black}{\downarrow\!\times4.36}}$
& $225_{\textcolor{green!65!black}{\downarrow\!\times1.41}}$ \\

\midrule

\textbf{Mean LOC Reduction ratio}
& ---
& $\mathbf{10.85\times}$
& $\mathbf{5.46\times}$
& $\mathbf{1.79\times}$ \\

\bottomrule
\end{tabular}
\end{adjustbox}
\end{wraptable}

To address \ref{RQ1}, we evaluate the implementation overhead involved in introducing LLM functionality into an existing codebase. Following prior studies~\cite{lmql, mtp2025jayanaka}, we measure the lines of code (LOC) added or modified for this integration. The counts include developer-authored prompts and annotations, including SemTexts and DSPy signature descriptions and field annotations. We compare PE, Base MTP, SemEng (MTP + SemTexts), and DSPy across the seven applications in Table~\ref{tab:benchmarks}. This measurement captures the size of the authored integration; RQ3 separately examines cognitive effort through an agent-based experiment.

Table~\ref{tab:loc} shows that SemEng achieves an average $5.46\times$ reduction in integration LOC relative to PE and requires fewer lines than DSPy across all seven applications. These comparisons use implementations whose prompts provide the same application requirements and task information, as described in the baseline setup. Adding SemTexts increases LOC relative to Base MTP, with the largest additions occurring in Service Center and Fact Verification. Even in these applications, the resulting SemEng implementations remain smaller than both PE and DSPy.

\paragraph{\textbf{Key Takeaway}}
\textbf{Lightweight Semantic Engineering integration:} Adding SemTexts supplies missing application intent to the base MTP implementation through concise annotations, without extensive code modifications. Even with these annotations, SemEng requires fewer integration lines than PE and DSPy.

\subsection{RQ2: Application Performance of Semantic Engineering}
\label{sec:eval-accuracy}

To address \ref{RQ2}, we compare SemEng with PE, Base MTP, and DSPy on standard reasoning and question-answering benchmarks and the seven application benchmarks, using the three models described in the experimental setup.

\paragraph{\textbf{Standard Benchmarks}}

\begin{wraptable}{r}{0.68\textwidth}
\centering
\caption{Exact-match accuracy on GSM8K and HotpotQA. Each benchmark uses a fixed sample of 300 examples shared across all models and approaches. Higher scores indicate better performance.}
\label{tab:general-accuracy}

\footnotesize
\renewcommand{\arraystretch}{1.18}
\setlength{\tabcolsep}{3pt}

\begin{adjustbox}{max width=\linewidth}
\begin{tabular}{@{}lcccc@{}}
\toprule

\textbf{Model}
& \textbf{PE}
& \textbf{DSPy}
& \textbf{Base MTP}
& \makecell[c]{\textbf{MTP +}\\ \textbf{SemTexts}} \\

\midrule

\multicolumn{5}{@{}l}{\textbf{GSM8K --- Exact Match (\%)}} \\

Opus 5.5
& 98.3 & 99.0 & 99.0 & 98.7 \\

DeepSeek V4.1 Flash
& 98.0 & 98.3 & 98.0 & 98.0 \\

Gemma 3 27B
& 95.7 & 95.0 & 94.3 & 95.7 \\

\midrule

\multicolumn{5}{@{}l}{\textbf{HotpotQA --- Exact Match (\%)}} \\

Opus 5.5
& 68.0 & 70.0 & 67.0 & 69.7 \\

DeepSeek V4.1 Flash
& 66.0 & 66.7 & 60.3 & 66.7 \\

Gemma 3 27B
& 57.3 & 54.7 & 52.0 & 57.0 \\

\bottomrule
\end{tabular}
\end{adjustbox}
\end{wraptable}
We evaluate fixed random samples of 300 GSM8K problems~\cite{cobbe2021gsm8k} and 300 HotpotQA questions~\cite{yang2018hotpotqa}, shared across all models and approaches. We report exact match of the final numerical answer for GSM8K and answer exact match in HotpotQA's distractor setting. Each configuration is evaluated once at temperature zero. Table~\ref{tab:general-accuracy} shows that SemEng achieves scores close to those of PE and DSPy on both standard benchmarks. The approaches show little separation on GSM8K, while adding SemTexts improves Base MTP's HotpotQA scores across all three models. 

\begin{figure*}[t]
    \centering
    \includegraphics[width=\linewidth]{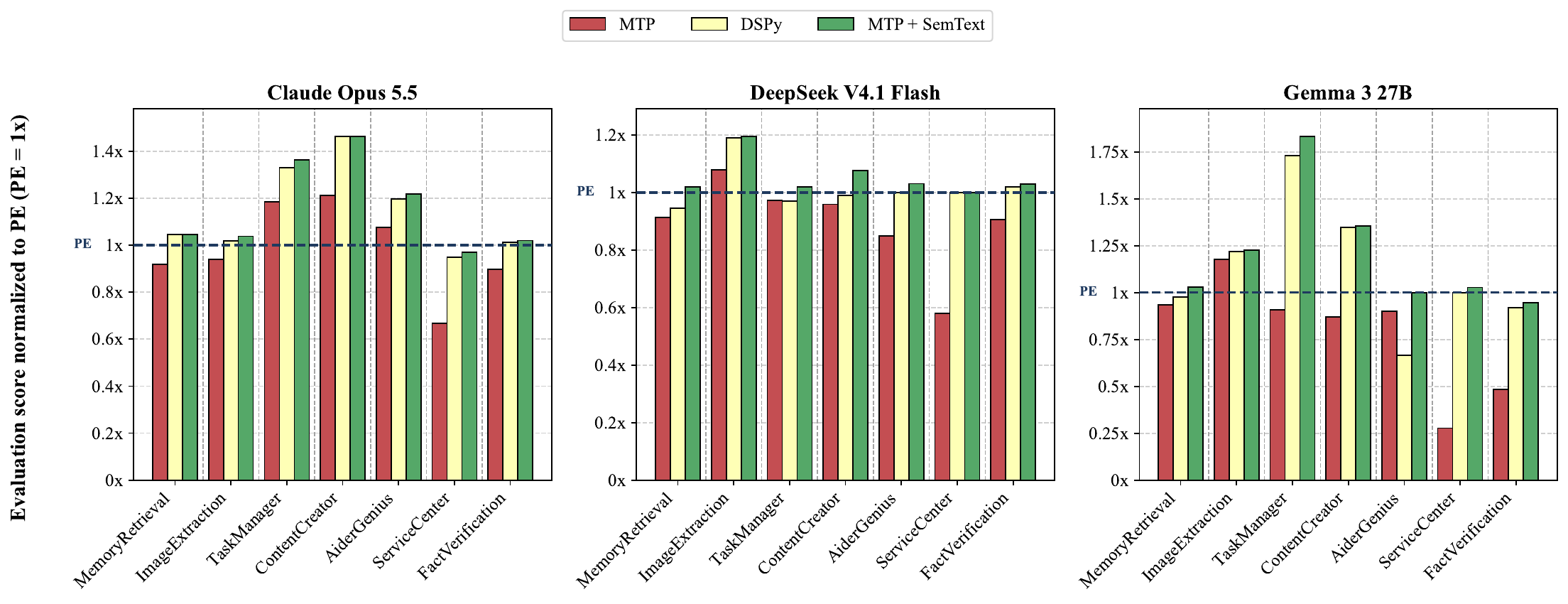}
    \caption{Application performance averaged over three repetitions per dataset instance. For each model and application, the mean scores of Base MTP, DSPy, and MTP + SemTexts are normalized by the corresponding mean PE score. The dashed line marks PE at 1.0; values above this line indicate higher mean scores than PE.}
    \Description{Three model-specific panels comparing Base MTP, DSPy, and MTP enhanced with SemTexts across seven application benchmarks. Scores are averaged across dataset instances and three repetitions per instance, then normalized by the corresponding mean PE score. A dashed line marks the PE reference at 1.0.}
    \label{fig:accuracy}
\end{figure*}

\paragraph{\textbf{Application Benchmarks}}
We evaluate the seven applications using the task-specific metrics in Table~\ref{tab:benchmarks}. For each model and integration approach, we run every dataset instance three times and report the mean score across all instances and repetitions. Figure~\ref{fig:accuracy} normalizes these mean scores by the corresponding mean PE score for the same model and application, with PE fixed at 1.0.

Adding SemTexts improves Base MTP's application performance, with clear gains on Task Manager and Service Center across all three models. These results show that explicit annotations can improve outcomes in the evaluated workflows involving tool use and agent coordination. Memory Retrieval shows smaller changes. Compared with PE, SemEng reaches or exceeds the reference score in most model--application combinations. On Gemma, for example, Task Manager and Content Creator both score above PE. SemEng also achieves scores similar to or higher than DSPy across the application suite. The exceptions relative to PE are Service Center on Opus and Fact Verification on Gemma, where SemEng remains below PE despite improving over Base MTP.

\paragraph{\textbf{Key Takeaways}}
\begin{enumerate}
    \item \textbf{SemTexts improve application behavior:} Adding SemTexts improves Base MTP's task performance, with clear gains in applications involving tool use and multi-agent coordination.

    \item \textbf{Lightweight integration with competitive performance:} SemEng performs as well as or better than PE and DSPy in most evaluated settings, with lower implementation overhead.
\end{enumerate}

\subsection{RQ3: Cognitive Effort of Semantic Enrichment}
\label{sec:eval-cognitive-effort}

To address \ref{RQ3}, we examine the cognitive effort required to express missing application intent using SemEng and PE, with both approaches aiming to reach the same performance target. RQ1 measures the amount of integration code and annotations in the final implementation. Prior work shows that reasoning-token counts predict human response times across reasoning tasks, linking token consumption to cognitive effort~\cite{devarda2025cost}. Motivated by this finding, we use coding-agent token consumption as an indirect proxy for the effort needed to determine where and how to express missing intent. We run all approaches with the same coding agent, model, requirements, and performance target.

\paragraph{\textbf{Experimental Setup}}
We conduct the experiment on Task Manager and Service Center using Claude Code with Opus 5.5. We construct the starting implementations from the applications evaluated in RQ1 and RQ2, retaining their existing integration code and orchestration. For MTP, we remove all SemTexts while retaining the application code. For PE, we remove the corresponding application requirements from the manually written prompts, retaining the information that base MTP conveys through code semantics and runtime inputs. Both starting implementations therefore provide the application model with the same task information. The coding agent must then express the same additional application requirements through prompts or SemTexts.

For each application, the coding agent receives the same behavioral requirements and development tests. We compare three approaches: \textbf{Prompt Engineering(Prompt Eng.)}, which supplies the missing intent through application prompts; \textbf{Full SemText Annotation}, which adds SemTexts to all eligible program entities; and \textbf{Semantic Engineering(SemEngng)}, which identifies intent not already conveyed by the code and selectively adds or refines SemTexts at the relevant entities.

For each application and approach, we repeat the experiment three times using the same starting implementation and initial instructions. In each repetition, the agent uses development tests to test and revise its prompts or SemTexts until it meets or exceeds the existing PE baseline's score on those tests. We report the mean and standard deviation of coding-agent token consumption across the three repetitions. Task Manager uses DeepSeek V4.1 Flash, with 30 queries available to the agent as development tests and 50 separate queries reserved for final tests. Service Center uses Gemma~3~27B, with 41 development tickets and 78 separate tickets reserved for final tests.

\begin{wrapfigure}{r}{0.5\textwidth}
\centering
\includegraphics[width=\linewidth]{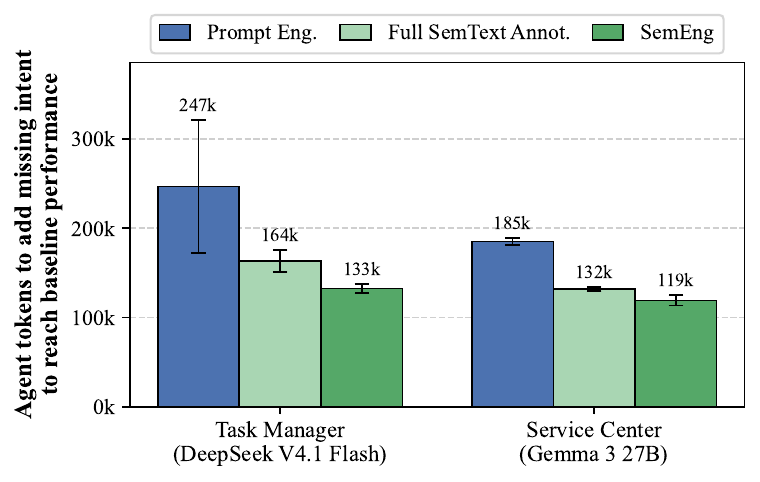}
\caption{Coding-agent tokens consumed to reach the PE baseline's score on the development tests. Bars show means over three repetitions, with error bars indicating one standard deviation. \emph{Prompt Eng.}: Prompt Engineering; \emph{Full Annot.}: full SemText annotation; \emph{SemEngng}: Semantic Engineering.}
\label{fig:cognitive-effort}
\end{wrapfigure}

\paragraph{\textbf{Results}}
Figure~\ref{fig:cognitive-effort} shows that both SemText approaches use fewer coding-agent tokens on average than PE to reach the PE baseline's score on the development tests. SemEng consumes the fewest tokens on both applications: PE uses approximately $1.86\times$ as many tokens as SemEng on Task Manager and $1.55\times$ as many on Service Center.

SemEng also reaches this target with fewer annotations than full SemText annotation: 7--11 versus 39 on Task Manager, and 11 versus 30--31 on Service Center. In these experiments, selectively annotating the entities associated with missing intent is sufficient to reach the baseline performance, with lower token consumption than annotating all eligible entities.

On the final tests, SemEng achieves scores close to those of the agent-authored PE implementations: 80.2\% versus 79.6\% on Task Manager and 91.5\% versus 92.7\% on Service Center. Thus, SemEng uses fewer coding-agent tokens to reach the development target while achieving similar observed performance to agent-authored PE on the final tests.

These results suggest that SemEng requires less cognitive effort than PE to express missing application intent and reach the same performance target. PE requires locating and revising the relevant prompt segments. SemEng allows the missing intent to be expressed through SemTexts at the relevant program entities, while MTP incorporates these additions into generated prompts. The lower token consumption is consistent with this structure reducing the effort needed to determine where and how to express the missing intent.

\paragraph{\textbf{Key Takeaways}}
\begin{enumerate}
    \item \textbf{Lower cognitive effort to add missing intent:} The results suggest that SemEng requires less cognitive effort than PE to add missing application intent and reach the same performance target.

    \item \textbf{Selective enrichment reduces effort:} The results suggest that adding SemTexts only where intent is missing reaches the same performance target with less cognitive effort than annotating every eligible entity.
\end{enumerate}

\subsection{RQ4: Understanding Where and Why Semantic Enrichment Helps}
\label{sec:eval-semantic-enrichment}

\begin{figure*}[!tp]
    \centering
    \includegraphics[width=\linewidth,height=0.78\textheight,keepaspectratio]{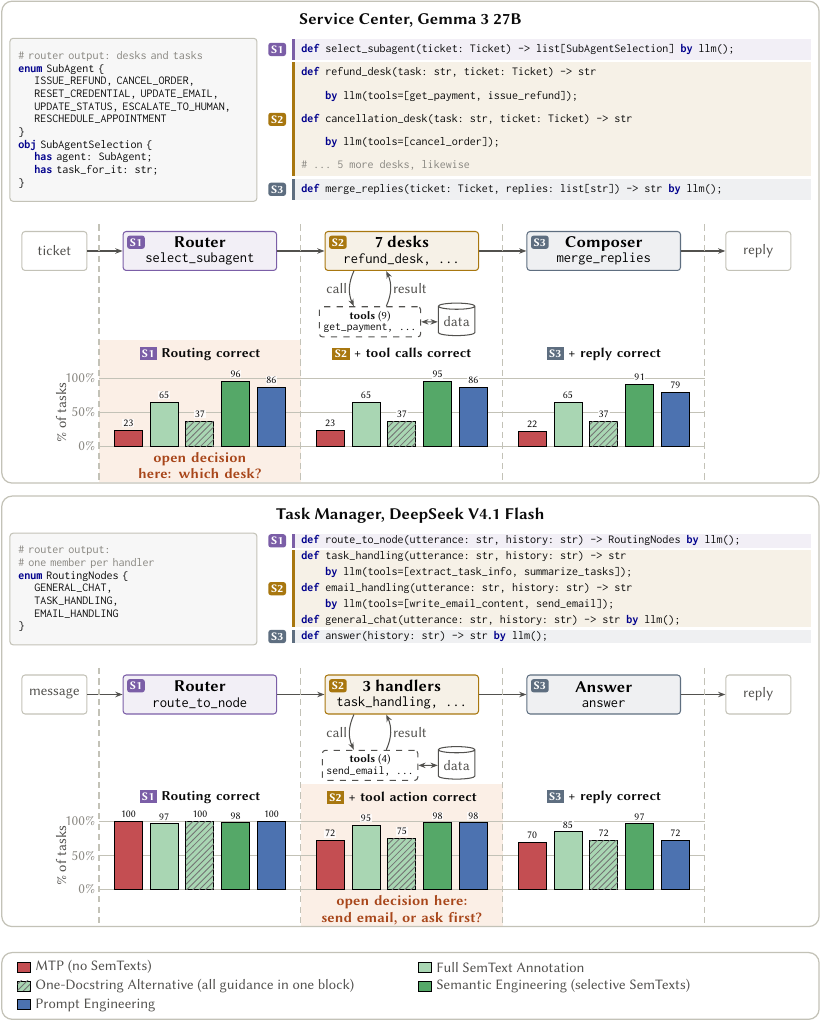}
    \caption{Application structure and cumulative task success for Service Center using Gemma~3~27B and Task Manager using DeepSeek V4.1 Flash. S1 counts tasks routed correctly. S2 counts tasks that pass both routing and desk or tool-action checks. S3 counts tasks that also pass the final-reply checks. Every percentage uses the full task set as its denominator. The selective Service Center configuration includes both routing and desk SemTexts but no composer SemText.}
    \label{fig:stage-code}

\end{figure*}

To address \ref{RQ4}, we examine when application requirements are not sufficiently expressed by code and where SemTexts should be added to supply the missing intent.

\paragraph{\textbf{Experimental Setup}}
We evaluate two benchmark applications from Table~\ref{tab:benchmarks}: Service Center with Gemma~3~27B and Task Manager with DeepSeek V4.1 Flash. For this ablation study, we select 60 of the 240 Service Center tickets and 60 of the 220 Task Manager queries. We evaluate all configurations on the same sample for each application, including the MTP baseline (\emph{base MTP}) and Prompt Engineering (PE) from the earlier experiments. We also evaluate three additional configurations: \emph{MTP + Docstrings}, which adds one docstring per relevant function, with enum-selection criteria collected in the routing function's docstring; \emph{Full SemText Annotation}, which adds SemTexts to all eligible program entities; and \emph{Semantic Engineering (SemEngng)}, which selectively adds SemTexts where application requirements are not sufficiently expressed by code. We repeat each evaluation three times and report mean success rates.

\begin{wrapfigure}{r}{0.5\textwidth}
    \centering
    \includegraphics[width=\linewidth]{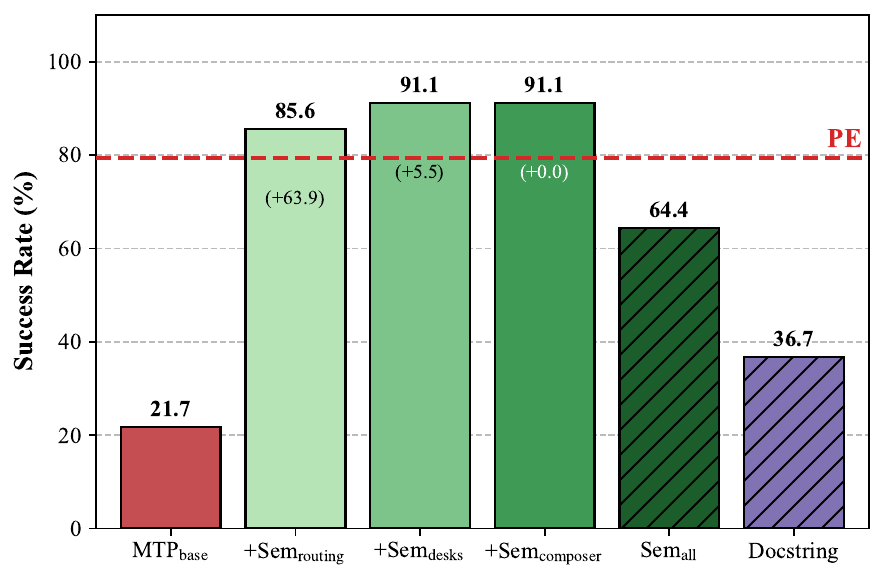}
    \caption{Final Service Center success using Gemma~3~27B. Routing, desk, and composer SemTexts are added cumulatively. Full annotation and docstrings are separate comparisons; the dashed line shows PE.}
    \Description{Final success is 21.9 percent for base MTP, 85.6 percent after adding routing annotations, 91.0 percent after adding desk annotations, and 91.2 percent after adding the composer annotation. Full annotation achieves 64.7 percent and function-level docstrings achieve 36.8 percent.}
    \label{fig:sc-ablation}
\end{wrapfigure}

Figure~\ref{fig:stage-code} shows the workflows and success rates at each stage for Service Center and Task Manager. For Service Center, Figure~\ref{fig:sc-ablation} shows how final task success changes as SemTexts are added incrementally, and Figure~\ref{fig:sc-semtexts} lists the SemTexts used.

\paragraph{\textbf{Service Center}}
Service Center resolves customer support tickets in three stages. First, the router, \texttt{select\_subagent}, selects the appropriate desks from seven available specialists. It returns a list of \texttt{SubAgentSelection} values, each pairing a specialist from the \texttt{SubAgent} enum with a delegated task. In the second stage, the selected desks carry out their assigned tasks using backend tools such as \texttt{get\_payment} and \texttt{issue\_refund}. Finally, \texttt{merge\_replies} combines their replies into one customer response. A ticket succeeds only when the backend reaches the expected state and the reply states the required facts.

We first compare the routing requirements with the code for \texttt{select\_subagent} and the \texttt{SubAgent} enum. We find missing selection criteria at both levels. The router's name and list return type do not specify one selection per currently requested action or the exclusion of background information. Within the enum, \texttt{UPDATE\_STATUS} can suggest changing status although its desk only reports it, while \texttt{CANCEL\_ORDER} leaves unclear that cancellation requests should still be routed when an order may have shipped. Similarly, \texttt{ESCALATE\_TO\_HUMAN} does not convey that escalation applies only when no other desk can handle the request and must be selected alone.

We express these missing requirements through $\mathrm{Sem}_{\mathrm{routing}}$: one SemText on the routing function and three on the relevant enum members, shown in lines 1--6 of Figure~\ref{fig:sc-semtexts}. In MTP + Docstrings, the routing requirements are collected in a single docstring on \texttt{select\_subagent}, including the selection criteria associated with individual \texttt{SubAgent} enum members. In Figure~\ref{fig:stage-code}, the percentage of all tickets routed correctly (S1) rises from 23\% with base MTP to 96\% with selective SemEngng, exceeding both PE and MTP + Docstrings. This result suggests that associating requirements with the relevant program entities is more effective than collecting them in a single function-level docstring.

We then compare the second-stage requirements with the code for the specialist desks. Their tool names and signatures identify the available operations, but two requirements concerning their replies remain missing. The status desk must state the status exactly as returned by its tool, and the escalation desk must tell the customer that a human agent will take over. We express these requirements through two function-associated SemTexts, $\mathrm{Sem}_{\mathrm{desks}}$, shown in lines 8--9 of Figure~\ref{fig:sc-semtexts}. With both $\mathrm{Sem}_{\mathrm{routing}}$ and $\mathrm{Sem}_{\mathrm{desks}}$, 95\% of all tickets are routed correctly and also pass the desk and tool-action checks (S2) in Figure~\ref{fig:stage-code}, exceeding base MTP and PE.

At the third stage, we compare the reply-composition requirement with the code for \texttt{merge\_rep}-\texttt{lies}. Its name, signature, and inputs already convey that it combines the specialist replies into one customer response, so we add no composer SemText. With only the routing and desk SemTexts, 91\% of all tickets pass the routing, desk and tool-action, and final-reply checks (S3) in Figure~\ref{fig:stage-code}. This final task success rate exceeds PE's 79\%, as well as the rates achieved by full SemText annotation and MTP + Docstrings. Thus, annotating all eligible entities or supplying guidance through function-level docstrings yields lower final success than selectively adding the six SemTexts.

Figure~\ref{fig:sc-ablation} then examines how final task success changes as SemTexts are added incrementally. Base MTP achieves 21.9\% final task success. Adding only the four routing SemTexts raises it to 85.6\%. Retaining those SemTexts and adding the two desk SemTexts raises final success to 91.0\%.

We also test an additional composer annotation, $\mathrm{Sem}_{\mathrm{composer}}$, shown on line 11 of Figure~\ref{fig:sc-semtexts}, which restates the composer's purpose. Adding it to the routing and desk annotations changes final success only from 91.0\% to 91.2\%. Thus, the largest improvement comes from supplying the missing routing criteria, followed by the two required reply facts; restating the composer's purpose provides little additional benefit.

\begin{figure}[!htbp]
\begin{jac}
sem select_subagent = "One selection per action the customer asks for now; most tickets need one."
                        "Ignore what the ticket mentions only as background, a cause, or a conditional plan.";
sem SubAgent.UPDATE_STATUS = "Tells the customer where an order is (its shipping status); changes nothing.";
sem SubAgent.CANCEL_ORDER = "Also when the order may already have shipped; the desk reports if it cannot.";
sem SubAgent.ESCALATE_TO_HUMAN = "Only needs no other desk can do (replacements, address changes,"
                                     "new appointments, complaints); never together with another desk.";

sem status_desk = "State the status word exactly as the tool returned it.";
sem escalation_desk = "Tell the customer that a human agent will take over the request.";

sem merge_replies = "Combine the desk replies into one reply to the customer for this ticket.";
\end{jac}
\caption{SemTexts used in the incremental Service Center study. Lines 1--6 supply routing criteria through annotations on the routing function and relevant enum members. Lines 8--9 add required reply facts at the specialist desks. Line 11 restates the composer's purpose. The selective SemEngng configuration includes the routing and desk annotations.}
\Description{Four routing annotations specify selection criteria, two desk annotations specify required reply facts, and one composer annotation describes combining specialist replies.}
\label{fig:sc-semtexts}
\end{figure}

\paragraph{\textbf{Task Manager}}
Task Manager illustrates a different location for missing intent. Its routing enum distinguishes \texttt{GENERAL\_CHAT}, \texttt{TASK\_HANDLING}, and \texttt{EMAIL\_HANDLING}. In Figure~\ref{fig:stage-code}, 100\% of all tasks are routed correctly with base MTP, compared with 98\% under the selective SemEngng configuration (S1). The selective configuration adds no routing SemTexts.

The missing requirement concerns email handling: the application must obtain the user's approval before sending an email, but the availability of \texttt{send\_email} does not express this condition. SemEngng supplies that requirement through a SemText while leaving routing and reply composition unannotated. Under this selective configuration, 98\% of all tasks pass both routing and the required task or email-handling checks (S2), compared with 72\% under base MTP.

Finally, 97\% of all tasks pass the routing, task or email-handling, and final-reply checks (S3) under the selective configuration, compared with 70\% under base MTP. The improvement therefore occurs in handling requests and completing tasks, while routing success remains similar. Selective enrichment also exceeds full annotation's 85\% final success, again showing that adding SemTexts to more entities does not necessarily improve behavior.

\paragraph{\textbf{Key Takeaways}}
\begin{enumerate}
    \item \textbf{Add SemTexts for selection criteria and policies missing from code:} Making these requirements explicit at the relevant program entities can improve task success.

    \item \textbf{Where intent is expressed matters:} Associating requirements with the relevant program entities through SemTexts can be more effective than combining them in a single function-level docstring.

    \item \textbf{Selective enrichment can outperform full annotation:} Annotating every eligible entity does not necessarily improve task success. Additional annotations may provide little benefit or reduce performance.
\end{enumerate}

\section{Related Work}

A wide range of approaches integrate LLMs into software through code generation~\cite{pythoness, code_gen} and runtime model calls~\cite{dspy, mtp2025jayanaka, lmql, sglang}. Our work focuses on runtime model calls and the application intent that code alone does not express.

LMQL~\cite{lmql} supports scripted prompts and constrained generation, while SGLang~\cite{sglang} focuses on efficient execution of structured model programs. APPL~\cite{appl2025} embeds prompts in Python functions, and POML~\cite{poml2025} organizes prompt content through reusable markup components. These systems structure prompt construction and execution; SemEng focuses on expressing additional intent through annotations associated with program entities. DSPy~\cite{dspy} expresses tasks through signatures, including task instructions and field descriptions, and supports automatic prompt optimization. MTP~\cite{mtp2025jayanaka} generates prompts from the semantics conveyed by program structure and runtime inputs. We build on MTP to let developers selectively supply application requirements not explicit in the code while retaining automated prompt construction.

TypeChat~\cite{microsoft_typechat} uses types and comments to guide structured responses, BAML~\cite{baml_descriptions} attaches descriptions to fields and enum values, and NVIDIA Object-Oriented Agents (NOOA)~\cite{nooa2026} uses docstrings as agent prompts. SemEng provides a dedicated annotation mechanism, separate from documentation and schema metadata, that binds intent to program entities. Declaration scopes determine which annotations apply to a call, and these associations are preserved in generated prompts. The natural-language content remains interpreted by the model.

GEPA~\cite{gepa} improves prompts through reflection on execution traces and feedback. This optimization is complementary to SemEng: GEPA searches for effective prompt content, while SemEng defines how additional intent is associated with code entities and selected for model calls.

Automatic docstring generation~\cite{poudel2024documintdocstringgenerationpython, yang2025docagentmultiagentautomatedcode} aims to document existing code. SemEng focuses on expressing additional application requirements. Our evaluation compares how docstrings and SemTexts associate that information with program entities during prompt construction.

\section{Conclusion}
This paper presents Semantic Engineering, a methodology for identifying missing application intent and selectively adding and refining model guidance. We realize it through SemTexts, which bind guidance to program entities, with compiler and runtime support for scoped selection during automatic prompt construction. Our evaluation shows a $5.46\times$ reduction in integration LOC relative to Prompt Engineering while matching or exceeding its task performance in most model--application settings, reaching up to $1.83\times$ its task performance. The coding-agent study shows lower token consumption to reach the same development-test targets, while the case studies show that selective enrichment can outperform full annotation and function-level docstrings. These findings support expressing missing intent at relevant program entities and refining it through testing. Future work will explore automated SemText suggestion and refinement.

\bibliographystyle{ACM-Reference-Format}
\bibliography{references}

@article{mtp2025jayanaka,
author = {Dantanarayana, Jayanaka L. and Kang, Yiping and Sivasothynathan, Kugesan and Clarke, Christopher and Li, Baichuan and Kashmira, Savini and Flautner, Krisztian and Tang, Lingjia and Mars, Jason},
title = {MTP: A Meaning-Typed Language Abstraction for AI-Integrated Programming},
year = {2025},
issue_date = {October 2025},
publisher = {Association for Computing Machinery},
address = {New York, NY, USA},
volume = {9},
number = {OOPSLA2},
url = {https://doi.org/10.1145/3763092},
doi = {10.1145/3763092},
journal = {Proc. ACM Program. Lang.},
month = oct,
articleno = {314},
numpages = {29}
}

@misc{dspy,
      title={DSPy: Compiling Declarative Language Model Calls into Self-Improving Pipelines}, 
      author={Omar Khattab and Arnav Singhvi and Paridhi Maheshwari and Zhiyuan Zhang and Keshav Santhanam and Sri Vardhamanan and Saiful Haq and Ashutosh Sharma and Thomas T. Joshi and Hanna Moazam and Heather Miller and Matei Zaharia and Christopher Potts},
      year={2023},
      eprint={2310.03714},
      archivePrefix={arXiv},
      primaryClass={cs.CL},
      url={https://arxiv.org/abs/2310.03714}, 
}

@article{lmql,
author = {Beurer-Kellner, Luca and Fischer, Marc and Vechev, Martin},
title = {Prompting Is Programming: A Query Language for Large Language Models},
year = {2023},
issue_date = {June 2023},
publisher = {Association for Computing Machinery},
address = {New York, NY, USA},
volume = {7},
number = {PLDI},
url = {https://doi.org/10.1145/3591300},
doi = {10.1145/3591300},
journal = {Proc. ACM Program. Lang.},
month = jun,
articleno = {186},
numpages = {24}
}

@article{cobbe2021gsm8k,
  title={Training Verifiers to Solve Math Word Problems},
  author={Cobbe, Karl and Kosaraju, Vineet and Bavarian, Mohammad and Chen, Mark and Jun, Heewoo and Kaiser, Lukasz and Plappert, Matthias and Tworek, Jerry and Hilton, Jacob and Nakano, Reiichiro and Hesse, Christopher and Schulman, John},
  journal={arXiv preprint arXiv:2110.14168},
  year={2021}
}

@inproceedings{yang2018hotpotqa,
  title={{HotpotQA}: A Dataset for Diverse, Explainable Multi-hop Question Answering},
  author={Yang, Zhilin and Qi, Peng and Zhang, Saizheng and Bengio, Yoshua and Cohen, William W. and Salakhutdinov, Ruslan and Manning, Christopher D.},
  booktitle={Conference on Empirical Methods in Natural Language Processing ({EMNLP})},
  year={2018}
}

@misc{laion-image,
      title={LAION-400M: Open Dataset of CLIP-Filtered 400 Million Image-Text Pairs}, 
      author={Christoph Schuhmann and Richard Vencu and Romain Beaumont and Robert Kaczmarczyk and Clayton Mullis and Aarush Katta and Theo Coombes and Jenia Jitsev and Aran Komatsuzaki},
      year={2021},
      eprint={2111.02114},
      archivePrefix={arXiv},
      primaryClass={cs.CV},
      url={https://arxiv.org/abs/2111.02114}, 
}

@article{chia2023instructeval,
      title={INSTRUCTEVAL: Towards Holistic Evaluation of Instruction-Tuned Large Language Models}, 
      author={Yew Ken Chia and Pengfei Hong and Lidong Bing and Soujanya Poria},
      journal={arXiv preprint arXiv:2306.04757},
      year={2023}
}

@inproceedings{
    jimenez2024swebench,
    title={{SWE}-bench: Can Language Models Resolve Real-world Github Issues?},
    author={Carlos E Jimenez and John Yang and Alexander Wettig and Shunyu Yao and Kexin Pei and Ofir Press and Karthik R Narasimhan},
    booktitle={The Twelfth International Conference on Learning Representations},
    year={2024},
    url={https://openreview.net/forum?id=VTF8yNQM66}
}

@ARTICLE{jac_cal,
  author={Mars, Jason and Kang, Yiping and Daynauth, Roland and Li, Baichuan and Mahendra, Ashish and Flautner, Krisztian and Tang, Lingjia},
  journal={IEEE Computer Architecture Letters}, 
  title={The Jaseci Programming Paradigm and Runtime Stack: Building Scale-Out Production Applications Easy and Fast}, 
  year={2023},
  volume={22},
  number={2},
  pages={101-104},
  doi={10.1109/LCA.2023.3274038}}

@misc{mars2025extendingdataspatialsemantics,
      title={Extending Data Spatial Semantics for Scale Agnostic Programming}, 
      author={Jason Mars},
      year={2025},
      eprint={2504.03109},
      archivePrefix={arXiv},
      primaryClass={cs.PL},
      url={https://arxiv.org/abs/2504.03109}, 
}

@misc{mars2025objectspatialprogramming,
      title={Object-Spatial Programming}, 
      author={Jason Mars},
      year={2025},
      eprint={2503.15812},
      archivePrefix={arXiv},
      primaryClass={cs.PL},
      url={https://arxiv.org/abs/2503.15812}, 
}

@misc{weber2024largelanguagemodelssoftware,
      title={Large Language Models as Software Components: A Taxonomy for LLM-Integrated Applications}, 
      author={Irene Weber},
      year={2024},
      eprint={2406.10300},
      archivePrefix={arXiv},
      primaryClass={cs.SE},
      url={https://arxiv.org/abs/2406.10300}, 
}

@article{scallop,
author = {Li, Ziyang and Huang, Jiani and Naik, Mayur},
title = {Scallop: A Language for Neurosymbolic Programming},
year = {2023},
issue_date = {June 2023},
publisher = {Association for Computing Machinery},
address = {New York, NY, USA},
volume = {7},
number = {PLDI},
url = {https://doi.org/10.1145/3591280},
doi = {10.1145/3591280},
journal = {Proc. ACM Program. Lang.},
month = jun,
articleno = {166},
numpages = {25}
}

@misc{sharma2025promptpexautomatictestgeneration,
      title={PromptPex: Automatic Test Generation for Language Model Prompts}, 
      author={Reshabh K Sharma and Jonathan De Halleux and Shraddha Barke and Benjamin Zorn},
      year={2025},
      eprint={2503.05070},
      archivePrefix={arXiv},
      primaryClass={cs.SE},
      url={https://arxiv.org/abs/2503.05070}, 
}

@misc{sahoo2025systematicsurveypromptengineering,
      title={A Systematic Survey of Prompt Engineering in Large Language Models: Techniques and Applications}, 
      author={Pranab Sahoo and Ayush Kumar Singh and Sriparna Saha and Vinija Jain and Samrat Mondal and Aman Chadha},
      year={2025},
      eprint={2402.07927},
      archivePrefix={arXiv},
      primaryClass={cs.AI},
      url={https://arxiv.org/abs/2402.07927}, 
}

@article{code_gen,
author = {M\"{u}ndler, Niels and He, Jingxuan and Wang, Hao and Sen, Koushik and Song, Dawn and Vechev, Martin},
title = {Type-Constrained Code Generation with Language Models},
year = {2025},
issue_date = {June 2025},
publisher = {Association for Computing Machinery},
address = {New York, NY, USA},
volume = {9},
number = {PLDI},
url = {https://doi.org/10.1145/3729274},
doi = {10.1145/3729274},
journal = {Proc. ACM Program. Lang.},
month = jun,
articleno = {171},
numpages = {26}
}

@misc{sglang,
      title={SGLang: Efficient Execution of Structured Language Model Programs}, 
      author={Lianmin Zheng and Liangsheng Yin and Zhiqiang Xie and Chuyue Sun and Jeff Huang and Cody Hao Yu and Shiyi Cao and Christos Kozyrakis and Ion Stoica and Joseph E. Gonzalez and Clark Barrett and Ying Sheng},
      year={2024},
      eprint={2312.07104},
      archivePrefix={arXiv},
      primaryClass={cs.AI},
      url={https://arxiv.org/abs/2312.07104}, 
}

@misc{gepa,
      title={GEPA: Reflective Prompt Evolution Can Outperform Reinforcement Learning}, 
      author={Lakshya A Agrawal and Shangyin Tan and Dilara Soylu and Noah Ziems and Rishi Khare and Krista Opsahl-Ong and Arnav Singhvi and Herumb Shandilya and Michael J Ryan and Meng Jiang and Christopher Potts and Koushik Sen and Alexandros G. Dimakis and Ion Stoica and Dan Klein and Matei Zaharia and Omar Khattab},
      year={2025},
      eprint={2507.19457},
      archivePrefix={arXiv},
      primaryClass={cs.CL},
      url={https://arxiv.org/abs/2507.19457}, 
}

@misc{microsoft_typechat,
  author = {Microsoft},
  title = {TypeChat: Helps get well-typed responses from language models to build pragmatic natural language interfaces},
  howpublished = {\url{https://microsoft.github.io/TypeChat}},
  note = {Accessed: 2025-08-21}
}

@misc{pythoness,
      title={Effective LLM-Driven Code Generation with Pythoness}, 
      author={Kyla H. Levin and Kyle Gwilt and Emery D. Berger and Stephen N. Freund},
      year={2025},
      eprint={2501.02138},
      archivePrefix={arXiv},
      primaryClass={cs.PL},
      url={https://arxiv.org/abs/2501.02138}, 
}

@misc{docstrings,
      title={DocuMint: Docstring Generation for Python using Small Language Models}, 
      author={Bibek Poudel and Adam Cook and Sekou Traore and Shelah Ameli},
      year={2024},
      eprint={2405.10243},
      archivePrefix={arXiv},
      primaryClass={cs.SE},
      url={https://arxiv.org/abs/2405.10243}, 
}

@misc{yang2025docagentmultiagentautomatedcode,
      title={DocAgent: A Multi-Agent System for Automated Code Documentation Generation}, 
      author={Dayu Yang and Antoine Simoulin and Xin Qian and Xiaoyi Liu and Yuwei Cao and Zhaopu Teng and Grey Yang},
      year={2025},
      eprint={2504.08725},
      archivePrefix={arXiv},
      primaryClass={cs.SE},
      url={https://arxiv.org/abs/2504.08725}, 
}

@misc{poudel2024documintdocstringgenerationpython,
      title={DocuMint: Docstring Generation for Python using Small Language Models}, 
      author={Bibek Poudel and Adam Cook and Sekou Traore and Shelah Ameli},
      year={2024},
      eprint={2405.10243},
      archivePrefix={arXiv},
      primaryClass={cs.SE},
      url={https://arxiv.org/abs/2405.10243}, 
}

@misc{nooa2026furgale,
  author        = {Furgale, Paul and Klingler, Severin and Nolan, James and Staats, Matt and Di Lorenzo, Gaia and Martinez Abad, Elisa and Sch{\"u}ller, Christian and Dinu, Razvan and Devoto, Alessio and Berard, Pascal and Kaplun, Gal and Sarafian, Elad and Roveri, Riccardo and Derczynski, Leon and Cabral, Ricardo Silveira},
  title         = {{NVIDIA-labs OO Agents}: Native Python Object-Oriented Agents},
  year          = {2026},
  eprint        = {2607.20709},
  archivePrefix = {arXiv},
  primaryClass  = {cs.AI},
  doi           = {10.48550/arXiv.2607.20709},
  url           = {https://arxiv.org/abs/2607.20709}
}

@misc{typechat2023hejlsberg,
  author = {Hejlsberg, Anders and Lucco, Steve and Rosenwasser, Daniel and Boggan, Pierce and Madan, Umesh and Hopcroft, Mike and Chandrasekaran, Gayathri},
  title  = {Introducing {TypeChat}},
  year   = {2023},
  month  = jul,
  url    = {https://microsoft.github.io/TypeChat/blog/introducing-typechat/},
  note   = {Microsoft TypeChat blog, July 20, 2023}
}

@misc{pydantic2026schema,
  author = {{Pydantic}},
  title  = {{JSON Schema}},
  year   = {2026},
  url    = {https://pydantic.dev/docs/validation/latest/concepts/json_schema/},
  note   = {Pydantic documentation, accessed October 8, 2026}
}

@misc{dspy2026fieldsource,
  author = {{DSPy Contributors}},
  title  = {{DSPy} Source Code: {dspy/signatures/field.py}},
  year   = {2026},
  url    = {https://github.com/stanfordnlp/dspy/blob/main/dspy/signatures/field.py},
  note   = {Source code, accessed October 8, 2026}
}

@inproceedings{appl2025,
  title     = {{APPL}: A Prompt Programming Language for Harmonious Integration of Programs and Large Language Model Prompts},
  author    = {Dong, Honghua and Su, Qidong and Gao, Yubo and Li, Zhaoyu and Ruan, Yangjun and Pekhimenko, Gennady and Maddison, Chris J. and Si, Xujie},
  booktitle = {Proceedings of the 63rd Annual Meeting of the Association for Computational Linguistics (Volume 1: Long Papers)},
  year      = {2025},
  pages     = {1243--1266},
  publisher = {Association for Computational Linguistics},
  doi       = {10.18653/v1/2025.acl-long.63},
  url       = {https://aclanthology.org/2025.acl-long.63/}
}

@misc{poml2025,
  title         = {Prompt Orchestration Markup Language},
  author        = {Zhang, Yuge and Chen, Nan and Xu, Jiahang and Yang, Yuqing},
  year          = {2025},
  eprint        = {2508.13948},
  archivePrefix = {arXiv},
  primaryClass  = {cs.HC},
  url           = {https://arxiv.org/abs/2508.13948}
}

@misc{baml_descriptions,
  author = {{BoundaryML}},
  title  = {{BAML} Documentation: @description / @@description},
  url    = {https://docs.boundaryml.com/ref/attributes/description},
  note   = {Accessed October 14, 2026}
}

@misc{nooa2026,
  title         = {{NVIDIA}-labs {OO} Agents: Native {Python} Object-Oriented Agents},
  author        = {Furgale, Paul and Klingler, Severin and Nolan, James and Staats, Matt and Di Lorenzo, Gaia and Martinez Abad, Elisa and Sch{\"u}ller, Christian and Dinu, Razvan and Devoto, Alessio and Berard, Pascal and Kaplun, Gal and Sarafian, Elad and Roveri, Riccardo and Derczynski, Leon and Silveira Cabral, Ricardo},
  year          = {2026},
  eprint        = {2607.20709},
  archivePrefix = {arXiv},
  primaryClass  = {cs.AI},
  url           = {https://arxiv.org/abs/2607.20709}
}

@inproceedings{theologitis-etal-2026-claimdb,
  title     = {{C}laim{DB}: A Fact Verification Benchmark over Large Structured Data},
  author    = {Theologitis, Michael and Dammu, Preetam Prabhu Srikar and Shah, Chirag and Suciu, Dan},
  booktitle = {Proceedings of the 64th Annual Meeting of the {A}ssociation for {C}omputational {L}inguistics (Volume 1: Long Papers)},
  month     = jul,
  year      = {2026},
  address   = {San Diego, California, United States},
  publisher = {Association for Computational Linguistics},
  pages     = {34428--34451},
  doi       = {10.18653/v1/2026.acl-long.1589},
  url       = {https://aclanthology.org/2026.acl-long.1589/}
}

@inproceedings{li2023bird,
  title     = {Can {LLM} Already Serve as a Database Interface? A {BI}g Bench for Large-Scale Database Grounded Text-to-{SQL}s},
  author    = {Li, Jinyang and Hui, Binyuan and Qu, Ge and Yang, Jiaxi and Li, Binhua and Li, Bowen and Wang, Bailin and Qin, Bowen and Geng, Ruiying and Huo, Nan and Zhou, Xuanhe and Ma, Chenhao and Li, Guoliang and Chang, Kevin C.-C. and Huang, Fei and Cheng, Reynold and Li, Yongbin},
  booktitle = {Advances in Neural Information Processing Systems (NeurIPS)},
  volume    = {36},
  publisher = {Curran Associates, Inc.},
  year      = {2023}
}

@misc{yao2024tau,
  title         = {$\tau$-bench: A Benchmark for Tool-Agent-User Interaction in Real-World Domains},
  author        = {Yao, Shunyu and Shinn, Noah and Razavi, Pedram and Narasimhan, Karthik},
  year          = {2024},
  eprint        = {2406.12045},
  archivePrefix = {arXiv},
  primaryClass  = {cs.AI},
  url           = {https://arxiv.org/abs/2406.12045}
}

@article{devarda2025cost,
  title   = {The cost of thinking is similar between large
             reasoning models and humans},
  author  = {de Varda, Andrea Gregor and D'Elia, Ferdinando Pio
             and Kean, Hope and Lampinen, Andrew
             and Fedorenko, Evelina},
  journal = {Proceedings of the National Academy of Sciences},
  year    = {2025},
  volume  = {122},
  number  = {47},
  pages   = {e2520077122},
  doi     = {10.1073/pnas.2520077122}
}

\end{document}